\documentclass[12pt,a4paper]{article}
\usepackage{amsmath}
\usepackage[font={footnotesize,it}]{caption}
\usepackage[top=1.2in, bottom=1.2in, left=1in, right=1in]{geometry}

\DeclareCaptionStyle{italic}[justification=centering]{labelfont={bf},textfont={it},labelsep=colon}
 \usepackage[title]{appendix} 
 
\usepackage{graphicx,psfrag,epsf,textcomp}
\usepackage{enumerate, dsfont}
\usepackage[round]{natbib}
\usepackage{xcolor} 
\usepackage[hyphens]{url}

\usepackage{breakurl}
\usepackage{booktabs, subfig, bm, paralist,mathpazo,tikz,todonotes,longtable,microtype}
\usepackage{color,soul,amsfonts,setspace,mathrsfs,mathtools}
\usepackage{multirow,array}
\usepackage[cal=boondox]{mathalfa}
\usepackage[pdftex,colorlinks=true]{hyperref}
\definecolor{darkblue}{rgb}{0,0,.6}
\definecolor{a0}{rgb}{0.0, 0.5, 0.0}
\definecolor{bistre}{rgb}{0.24, 0.17, 0.12}
\definecolor{amethyst}{rgb}{0.6, 0.4, 0.8}
\definecolor{blue-violet}{rgb}{0.54, 0.17, 0.89}
\definecolor{Rcolor}{RGB}{150,160,190}
\definecolor{blush}{rgb}{0.87, 0.36, 0.51}
\definecolor{brightturquoise}{rgb}{0.03, 0.91, 0.87}
\definecolor{burntorange}{rgb}{0.8, 0.33, 0.0}
\hypersetup{citecolor=darkblue,linkcolor=darkblue,urlcolor=darkblue}

\newcommand{\blind}{0}

\newcommand{\I}{\mathcal{I}}

\newcommand{\E}{\textit{E}}

\graphicspath{{plots/}}
\DeclareMathOperator*{\argmin}{\arg\!\min}

\newcolumntype{L}[1]{>{\raggedright\let\newline\\\arraybackslash\hspace{0pt}}m{#1}}
\newcolumntype{C}[1]{>{\centering\let\newline\\\arraybackslash\hspace{0pt}}m{#1}}
\newcolumntype{R}[1]{>{\raggedleft\let\newline\\\arraybackslash\hspace{0pt}}m{#1}}

\usepackage[ruled,linesnumbered]{algorithm2e}
\newtheorem{assumption}{Assumption}
\newtheorem{definition}{Definition}
\newtheorem{remark}{Remark}

\newsavebox\CBox

\newcommand{\inp}[2]{\langle #1\, , #2 \rangle}
\date{\today}
\usepackage{orcidlink}

\def\abs#1{\left\lvert#1\right\rvert}

\begin{document}

\def\spacingset#1{\renewcommand{\baselinestretch}
{#1}\small\normalsize} \spacingset{1}

\if0\blind
{
  \title{\bf Forecasting Australian fertility by age, region, and birthplace}
  \author{
  Yang Yang \orcidlink{0000-0002-8323-1490} \\
  Department of Econometrics and Business Statistics \\
  Monash University  \\
  \\
  Han Lin Shang\orcidlink{0000-0003-1769-6430} \footnote{Postal address: Department of Actuarial Studies and Business Analytics, Level 7, 4 Eastern Road, Macquarie University, Sydney, NSW 2109, Australia; Email: hanlin.shang@mq.edu.au} \\
  Department of Actuarial Studies and Business Analytics \\
  Macquarie University \\
\\
  James Raymer \orcidlink{0000-0001-6588-8985} \\
  School of Demography \\
  The Australian National University
 }
 \date{\today}
  \maketitle
} \fi

\if1\blind
{
  \title{\bf Forecasting Australian fertility by age, region, and birthplace}
\author{}
\date{}
\maketitle
} \fi

\bigskip
\begin{abstract}
Fertility differentials by urban-rural residence and nativity of women in Australia significantly impact population composition at sub-national levels. We aim to provide consistent fertility forecasts for Australian women characterized by age, region, and birthplace. Age-specific fertility rates at the national and sub-national levels obtained from census data between 1981--2011 are jointly modeled and forecast by the grouped functional time series method. Forecasts for women of each region and birthplace are reconciled following the chosen hierarchies to ensure that results at various disaggregation levels consistently sum up to the respective national total. Coupling the region of residence disaggregation structure with the trace minimization reconciliation method produces the most accurate point and interval forecasts. In addition, age-specific fertility rates disaggregated by the birthplace of women show significant heterogeneity that supports the application of the grouped forecasting method.
\vspace{.1in}

\end{abstract}

\spacingset{1.47}

\newpage
\section{Introduction} \label{sec:1}

National and sub-national fertility patterns in terms of age, geographical distribution, and birthplace are of interest to demographers, epidemiologists, health care personnel, government planners, and decision-makers \citep[see, e.g.,][]{rayer2009, raftery2012, kalasa2021}. Future fertility is a crucial component of overall population size and age structure and, therefore, will affect national population accounts and fiscal sustainability \citep{Eastwood1999}. According to the mother's age group, age-specific fertility rates are ratios of the annual births per 1,000 of the female estimated resident population in that age group. As a widely used measure of the age pattern of fertility in a community, age-specific fertility rates are important demographic metrics that ought to be accurately forecasted in practice.

Parametric and nonparametric fertility modeling and forecasting methods have been proposed \citep[see, e.g.,][for comprehensive reviews]{BT08, Shang12}. Recently, \cite{BLM18} assessed the forecast accuracy of 20 commonly used fertility forecasting methods. In a separate study, \cite{GST21} reviewed various fertility projection approaches applied to European data. In the parametric school of thought, \cite{Rogers86, TBL+89, Bell92}, and \cite{KP00} suggested fitting parametric distribution, such as Gamma distribution, to annual age-specific fertility rates. Then, the number of ages is reduced to a relatively smaller number of Gamma curve parameters; forecast the Gamma curve parameters using time-series techniques to forecast age-specific fertility rates. In the nonparametric school of thought, \cite{BB87} and \cite{Bell88, Bell92, Bell97} used the principal component analysis to obtain forecasts of age-specific fertility rates with only a small number of orthonormal principal components. From the viewpoint of mean squared error, these principal components and their associated scores capture the primary mode of information in the original data matrix.

Many fertility modeling applications in the literature only consider aggregated fertility at the national level. The fertility rates observed at the national level can be further disaggregated by characteristics of Australian women, namely age, birthplace, and region of residence. The Australian Government's Center for Population publishes annual forecasts of fertility at the national, state, and territory levels following assumptions that recent trends of observed births will continue for all age groups in the coming years \citep[see,e.g.,][]{CP21, Mcdonald2020}. Recently, academics studied fertility rates in capital cities and regional areas in Australia \citep[see, e.g.,][]{Wilson17, WMT20} and fertility rates of immigrants \citep[see, e.g.,][]{BRV21}. To the best of our knowledge, however, a comprehensive study of age-specific fertility rates taking consideration of geographical region of residence and birthplace of Australian females is previously lacking. This paper aims to fill this gap by providing national and sub-national fertility forecasts following hierarchical structures disaggregated by geographical locations and birthplace. 

The birthplace information presents a critical aspect linking fertility with migration. For instance: in 2020, the Australian total fertility rate was 1.65 births per woman, 1.68 for Australian-born women, and 1.55 for overseas-born women. Thus, immigrant fertility had little impact on the national fertility rate. The total fertility rate for the major immigrant groups did not differ significantly from the national fertility rate: 1.68 for United Kingdom-born, 2.07 for New Zealand-born, 1.58 for India-born, and 1.12 for China-born women \citep{ABSbirth}. The lower rate for the China-born women was explained entirely by their lower fertility in the age range 15-24, at which ages many Chinese women included in the population are international students living in Australia temporarily and thus unlikely to give birth. With the fertility forecasts by birthplace, we can study the temporal change in fertility within a migrant group and compare fertility between different migrant groups.

The region of residence at birth registration reveals the geographical variability in the fertility experience of Australian women. As one of the most urbanized countries in the world, Australia has over 67\% of its total population living in eight capital cities on 30 September 2021 \citep{ABSpopulation}. Significant geographic inequalities between different areas of the country in terms of income, education levels, employment opportunities, and population structure result in the substantial spatial variation of fertility in Australia \citep{EG2018}. Modeling and forecasting age-specific fertility rates of various regions identify fertility differences between metropolitan areas and rural or remote regions of the country.

A difficulty in regional fertility forecasting is that the forecasts at the regional levels are constrained to be sum to the forecasts at the national level. This problem gives rise to forecast reconciliation. Forecast reconciliation occurs in national account balancing \citep{SW09}, in tourist demand \citep{HAA+11}, in mortality forecasting \citep{SH17, LT19}, in annuity price forecasting \citep{SH17b}, in solar energy forecasting \citep{YYS20}, in electricity load and demand forecasting \citep{JPP19, NLP+20, BTH20}, and in tourism demand forecasting \citep{KA19}, to name only a few. This paper contributes a nonparametric approach to producing local age-specific fertility forecasts that consistently add to the national total.

A characteristic of age-specific fertility rates is the smooth shape over ages each year. We incorporate smoothness into the modeling and forecasting techniques to improve the accuracy of short-term forecasts. Implementing a smoothing technique can better capture the underlying trend of fertility changes, mitigating the influence of missing values and measurement noise in many sub-national series. With these aims in mind, we consider the functional time series method of \cite{HU07} to forecast national and sub-national age-specific fertility rates. As a generalization of \citeauthor{LC92}'s \citeyearpar{LC92} method, the functional time series method views age-specific fertility rates from a functional perspective, where age is treated as a continuum. We apply three existing forecast reconciliation methods, namely bottom-up, optimal combination, and trace minimization (MinT) methods, to reconcile fertility forecasts and improve forecast accuracy. The bottom-up method involves forecasting each of the disaggregated series and then using simple aggregation to obtain forecasts for the aggregated series \citep{Kahn98}. The method works well where the bottom-level series have a strong signal-to-noise ratio. The optimal combination method obtains forecasts independently at all levels of the group structure. Then, linear regression is used with a least-squares estimator to combine and reconcile the forecasts optimally. The MinT method incorporates the information from a full covariance matrix of forecast errors in obtaining a set of coherent forecasts. By minimizing the mean squared error of the coherent forecasts across all levels of disaggregation, the method performs well on empirical data.

The remainder of this paper is structured as follows. In Section~\ref{sec:2}, we describe the data set --- Australian national and sub-national fertility rates. In Section~\ref{sec:3}, we introduce a functional time series forecasting method for producing point and interval forecasts and then revisit the reconciliation methods in Section~\ref{sec:4}. We evaluate and compare point and interval forecast accuracies between the independent and grouped time series forecasting methods in Sections~\ref{sec:5.1} and~\ref{sec:5.2}, respectively. We consider the averaged root mean squared forecast error for evaluating the point forecast accuracy. The reconciled forecast methods generally produce the most accurate overall point forecasts, particularly the optimal combination with trace minimization. For evaluating the interval forecast accuracy, we consider the mean interval score. We found that the base and reconciled forecasts provide similar forecast accuracy, although the latter one eases forecast interpretability. Conclusions are presented in Section~\ref{sec:7}, along with some reflections on how the methodology can be extended further.

\section{Australian fertility by age, region, and birthplace}\label{sec:2}

\subsection{Overview of data}\label{sec:2.1}

We study the Australian age-specific fertility between 1981 and 2011\footnote{With quinquennial births and population data between 1981 and 2016, age-specific fertility rates at five-year intervals for 1981--2011 can be computed according to \eqref{eq_1} in Section~\ref{sec:2.2}. Fertility rates after 2011 are not considered since the EPRs for the interval 2016--2021 (i.e., $\frac{1}{2}\left(E_{g,j,2016}(x_p)+E_{g,j,2021}(x_p)\right)$) are not available.}. Fertility data of Australian female populations, including immigrants of various origins, are computed from birth registrations and the quinquennial census records. The number of births is prepared by state and territory Registries of Births, Deaths, and Marriages and are based on information forms completed by the parent(s) of newborns \citep{ABSbirth}. The national census occurs every five years and we consider the estimated resident populations (ERPs) of women in age groups from 15 to 49 (i.e., 15--19, 20--24, $\ldots$, 45--49) \citep{Census21}. For any particular age group of women, the age-specific fertility rates are computed as the quotient of average newborns in five consecutive calendar years divided by the average women's ERP. Further details on computing age-specific fertility rates are provided in Section~\ref{sec:2.2}.

In this paper, Australian age-specific fertility rates are disaggregated by government administrative divisions and country of birthplace groupings of female residents surveyed in the census. The harmonized time series of administrative units and birthplace groupings are described in \citet[][pp. 1059-1060]{RB18} and \cite{RSG+18}, respectively. The harmonized geographic units were based on various Statistical Division geographies from the Australian Standard Geographical Classification. To make the Statistical Division geographic boundaries consistent over time, they used simple rules that either assumed the boundary changes were insubstantial (if the boundary change resulted in only a small amount of population change) or merged multiple geographic areas into single (larger) ones. The resulting geography represented 47 areas that could be aggregated into 11 regions, as shown in Table~\ref{tab:1}.
 
\begin{onehalfspace}
\begin{center}
\small
\begin{longtable}{@{}L{3cm}C{1.3cm}L{3cm}C{0.2cm}L{2cm}C{1.3cm}L{4.5cm}@{}}
\caption{A list of Australian administrative division codes and names.} \label{tab:1} \\
\toprule
Region  & Area Index & Area Name &  & Region  & Area Index & Area Name  \\ \midrule
\endfirsthead
Region  & Area Index & Area Name &  & Region  & Area Index & Area Name  \\ \midrule
\endhead
\midrule
\multicolumn{7}{r}{{Continued on next page}} \\ 
\endfoot
\endlastfoot
Sydney ($R_1$) & 1 & Sydney & &  remote Australia ($R_{11}$) & 9 & West NSW \\ \\ \cmidrule{1-3}
NSW Coast ($R_2$) & 2 & Hunter & & & 22 & South West Queensland \\ 
& 3 & Illawarra & & & 23 & Central West Queensland \\ 
& 4 & Mid-North Coast & & & 25 & Far North \\ \cmidrule{1-3}
Melbourne ($R_3$) & 10 & Melbourne & & & 26 & North West \\ \cmidrule{1-3} 
Country VIC ($R_4$) & 11 & Barwon & & & 31 & Eyre  \\ 
& 12 & Western District & & & 32 & Northern SA  \\
& 13 & Central Highlands & & & 38 & South Eastern WA \\
& 16 & Loddon \& Goulbourn & & & 39 & Central WA \\
& 18 & Gippsland & & & 40 & Pilbara \\ \cmidrule{1-3}
Brisbane ($R_5$) & 19 & Brisbane & & & 41 & Kimberley \\ \cmidrule{1-3}
Adelaide ($R_6$) & 27 & Adelaide & & & 44 & Mersey-Lyell \\ \cmidrule{1-3}
Perth ($R_7$) & 33 & Perth & & & 46 & Northern Territory \\ \cmidrule{1-3} \cmidrule{5-7}
Hobart ($R_8$) & 42 & Hobart & & & & \\ \cmidrule{1-3}
ACT ($R_9$) & 47 & Canberra & & & & \\ \cmidrule{1-3}
Regional Australia ($R_{10}$) & 5 & North West NSW \\
& 6 & Central West NSW & & & & \\
& 7 & Murrumbidgee & & & & \\
& 8 & Murray & & & & \\
& 14 & Wimmera & & & & \\
& 15 & Mallee & & & & \\
& 17 & Ovens-Murray & & & & \\
& 20 & Wide Bay-Burnett & & & & \\
& 21 & Darling Downs & & & & \\
& 24 & Mackay \& Northern & & & & \\
& 28 & Yorke & & & & \\
& 29 & Murray Lands & & & & \\
& 30 & South East & & & & \\
& 34 & South West & & & & \\
& 35 & Lower Southern WA & & & & \\
& 36 & Upper Southern WA & & & & \\
& 37 & Midlands & & & & \\
& 43 & Northern Tasmania & & & & \\
& 45 & Darwin & & & &  \\
\bottomrule
\end{longtable}
\end{center}
\end{onehalfspace}

Birthplace information of females connects fertility with migration, where we could single out which migrant group contributes the most to the total Australian fertility. Disaggregating the Australian age-specific fertility rates by the birthplace of women yields 19 Country of Birth (COB) series, as listed in Table~\ref{tab:2}. The 19 COB covers almost all populated regions worldwide, which can be further divided into ten geographical zones according to the considered countries' cultural and economic connections.

\begin{onehalfspace}
\centering
\tabcolsep 0.48in
\renewcommand{\arraystretch}{1}
\begin{longtable}{@{}L{6cm}C{1.5cm}L{5.5cm}@{}}
\caption{List of COB for Australian female populations.} \label{tab:2} \\
\toprule
World Zones  & COB & COB Name \\ \midrule
\endfirsthead
World Zones  & COB & COB Name \\ \midrule
\endhead
Oceania ($Z_1$) & 1 & Australia \\
& 2 &  New Zealand \\
& 3 & Other Oceania and Antarctica \\ \cmidrule{1-3}
North-West Europe ($Z_2$) & 4 &  United Kingdom \\
& 5 & Other North-West Europe \\ \cmidrule{1-3}
South-East Europe ($Z_3$) & 6 & South-East Europe \\  \cmidrule{1-3}
North Africa and the Middle East ($Z_4$) & 7 & North Africa and the Middle East \\ \cmidrule{1-3}
South-East Asia ($Z_5$) & 8 & Vietnam \\
& 9 & Philippines \\
& 10 & Malaysia \\
& 11 & Indonesia \\
& 12 & Other South-East Asia \\ \cmidrule{1-3}
North-East Asia ($Z_6$) & 13 & China (excludes SARs, Taiwan) \\
& 14 &  Other North-East Asia  \\ \cmidrule{1-3}
Southern and Central Asia ($Z_7$) & 15 & India \\
& 16 & Other Southern and Central Asia \\  \cmidrule{1-3}
North America ($Z_8$) & 17 & North America \\  \cmidrule{1-3}
South America ($Z_9$) & 18 & South America \\ \cmidrule{1-3}
 Sub-Saharan Africa ($Z_{10}$) & 19 & Sub-Saharan Africa \\
\bottomrule
\end{longtable}
\end{onehalfspace}

Across all levels of disaggregation, there are $(1+11+47)\times 19 = 1121$ series, as summarized in Table~\ref{tab:3}. This study considers administrative division and birthplace the two main disaggregation factors in grouped time series models.

\begin{table}[!htb]
\centering
\caption{An overview of disaggregation factors for Australian age-specific fertility rates}\label{tab:3}
\begin{tabular}{@{}lr@{}}
\toprule
Level & Number of series \\
\midrule
Australia & 1 \\
Region & 11 \\
Area & 47 \\
Birthplace & 19 \\
\bottomrule
\end{tabular}
\end{table}

\subsection{Nonparametric smoothing applied to age-specific fertility rate} \label{sec:2.2}

For any considered birthplace $j$ of administrative division $j$, let $E_{g \ast j,t}(x_p)$ denote the ERP of women belonging to the age group $x_p$ in year $t$, where $x$ denotes the age variable, $p$ denotes the number of age groups considered. Let $I_{g \ast j,t}(x_p)$ denote the number of infants born to these women during the same calendar year. The women population at risk is computed at the midpoint of five-year intervals for a given age group as $\frac{1}{2}\left[E_{g \ast j,t}(x_p) + E_{g \ast j,t+5}(x_p)\right]$. For consistency with the computed ERP exposure, the total number of infants born in five consecutive calendar years are also averaged as $\frac{1}{5} \sum_{k=0}^{4} I_{g \ast j,t+k}(x_p)$ \citep{RB18}. The age-specific fertility rate is then calculated as the quotient of the average of newborns divided by the women population at risk as
\begin{equation}
  y_{g \ast j,t}(x_p) = \frac{\frac{1}{5} \sum_{k=0}^{4} I_{g \ast j,t+k}(x_p)}{\frac{1}{2} \left[E_{g \ast j,t}(x_p) + E_{g \ast j,t+5}(x_p)\right]} = \frac{2\sum_{k=0}^{4} I_{g \ast j,t+k}(x_p)}{5\left[E_{g \ast j,t}(x_p) + E_{g \ast j,t+5}(x_p)\right]} \label{eq_1}.
\end{equation}
Following~\eqref{eq_1}, we label each computed age-specific fertility rate with years covered by quinquennial censuses, i.e., ``1981-86'' for the five-year interval 1981--1986, and so on, ``2006-11'' for the five-year interval 2006--2011.

The birth registrations provided by the ABS report the number of newborns for seven age groups (i.e., 15--19, 20--24, \ldots, 45--49) over 5-year intervals between 1981 and 2011. The sparseness in the data occurs for some birthplace groups in regional and remote Australia. We first fill in missing observations by linear interpolation for each birthplace to obtain 35 observations between ages 15 and 49 every year between 1981 and 2011. The obtained raw age-specific fertility rates constitute 31 curves contaminated with noise and interpolation errors. No census ever recorded any female born in Other North-East Asia ($\text{COB}_{14}$) living in Central West Queensland ($\text{Area}_{23}$). Hence, we ignore the sub-national series $\text{Area}_{23} \times \text{COB}_{14}$ in this study. 

To obtain smooth functions and deal with possibly missing values, we consider a weighted regression $B$-splines with a concavity constraint for preserving the shape of the fertility curves \citep[see also][]{HU07}. Let $y_t^{(j)}(x_i)$ be the fertility rates observed at the beginning of each year for calendar year $t=1,2,\dots,n$ at ages $(x_1, x_2, \dots, x_N)$. We assume that there is an underlying continuous and smooth function $f_t^{(j)}(x)$ such that
\begin{equation*}
y_t^{(j)}(x_i) = f_t^{(j)}(x_i) + \delta_t^{(j)}(x_i)\varepsilon_{t,i}^{(j)},\qquad i=1,2,\dots,p,
\end{equation*}
where $\{\varepsilon_{t,i}^{(j)}\}$ are independent and identically distributed random variables across $t$ and $i$ with a mean of zero and a unit variance, and $\delta_t^{(j)}(x_i)$ measures the variability associated with fertility at each age in year $t$ for the $j$\textsuperscript{th} population. Jointly, $\delta_t^{(j)}(x_i)\varepsilon_{t,i}^{(j)}$ represents the smoothing error. The linear interpolation and nonparametric smoothing are useful for transforming data from five-year age groups into single-year. It is desirable to work with yearly data by single age in many statistical models \citep[see also][]{LGS+11}. 

\subsection{Fertility distribution} \label{sec:2.3}

\begin{figure}[!htb]
\centering
\subfloat{\includegraphics[width = 3.5in]{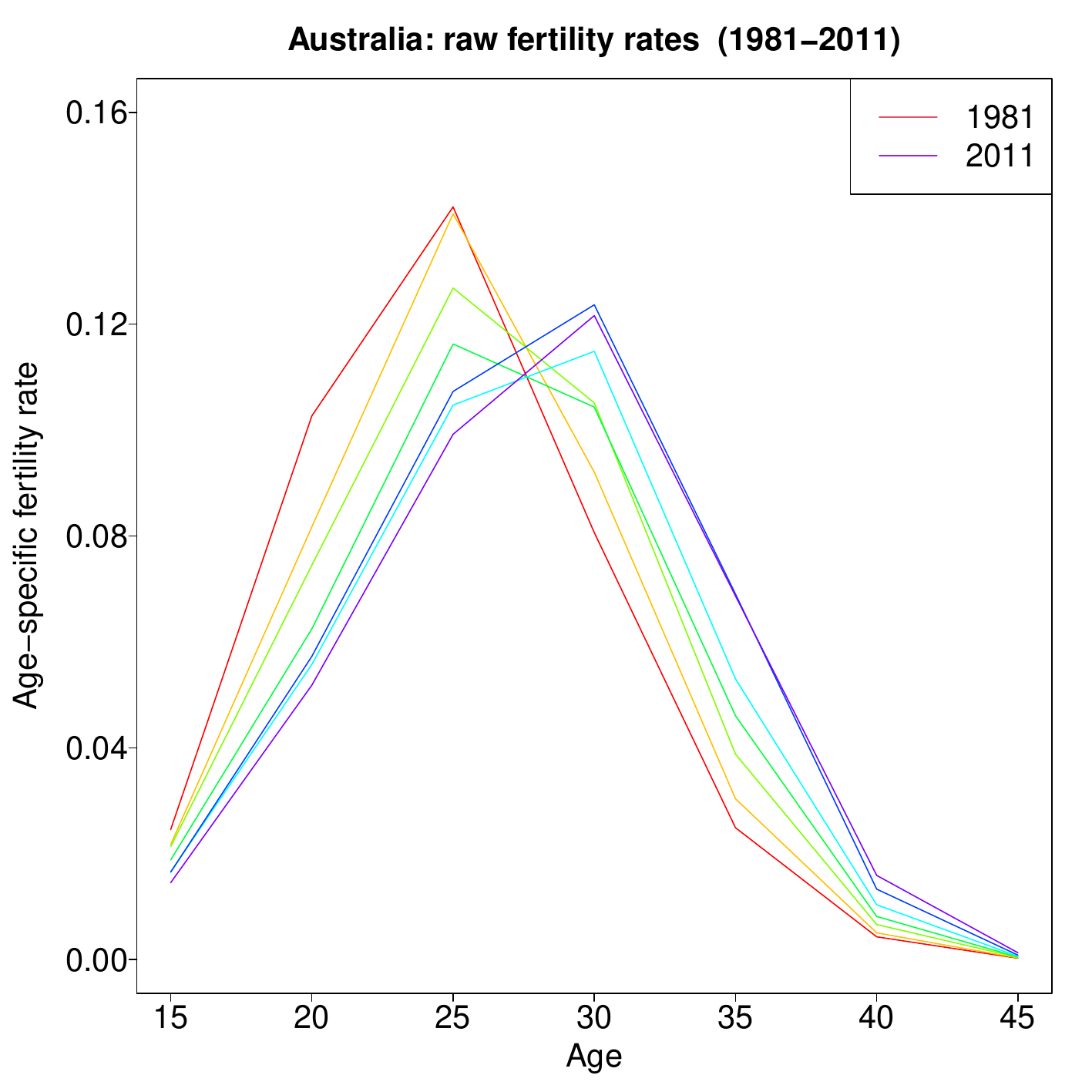}} 
\quad
\subfloat{\includegraphics[width = 3.5in]{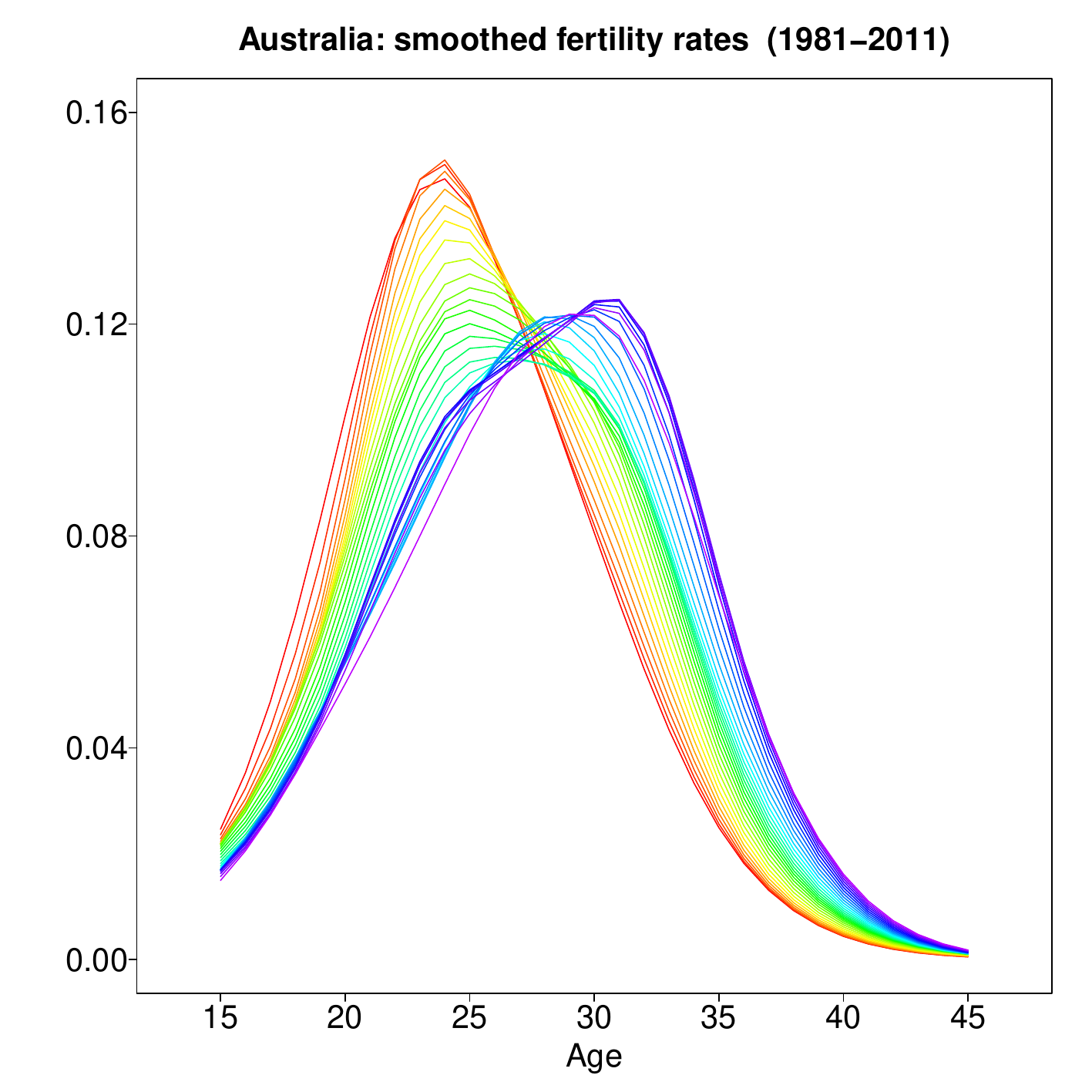}}
\caption{Australian age-specific fertility rates between 1981 and 2011} \label{fig:1}
\end{figure}

Figure~\ref{fig:1} presents rainbow plots of the Australian age-specific fertility rates in the period studied. The rainbow plot displays the distant past curves in red and the more recent curves in purple \citep{HS10}, indicating an overall reduction in the trend of Australian fertility rates. The figures show typical fertility curves for a developed country, with a postponement in childbearing \citep[see, e.g.,][]{BB12}. 

There are two most distinct patterns of the Australian age-specific fertility rates shown by the rainbow plots. In Figure~\ref{fig:1}, the left panel shows the observed fertility rates for all seven age groups obtained in the birth data. In contrast, the right panel presents the smoothed fertility curves for individual calendar years and single-year age groups after interpolation. 
\begin{inparaenum}
\item[1)] The overall distribution of Australian age-specific fertility rates appears to experience a rightward shift during the period. The peak fertility ages increase from 20--24 in 1981 to 30--35 in 2011. This decline in fertility of younger women is mainly caused by the postponement of childbearing by Australian females. In contrast, the increase in fertility of older women is due to the partial recovery of the previously postponed childbearing \citep{Lattimore2008}. By comparing the fertility curves of 1981 and 2011, the decline in age-specific fertility at younger ages (the area between the red and purple curves on the left side of the distribution) exceeds the growth of fertility rates of older women (the area between the red and purple curves on the right side of the distribution). 
\item[2)] Another feature of Australian age-specific fertility curves is the decline of the peak fertility rates over time. This decline in the height of age-specific fertility distribution is expected to negatively affect the average number of births an Australian woman would bear during her lifetime (i.e., the total fertility rate) \citep[see, e.g.,][]{Bongaarts1998}.
\end{inparaenum}

\begin{figure}[!htb]
\centering
\subfloat{\includegraphics[width = 3.5in]{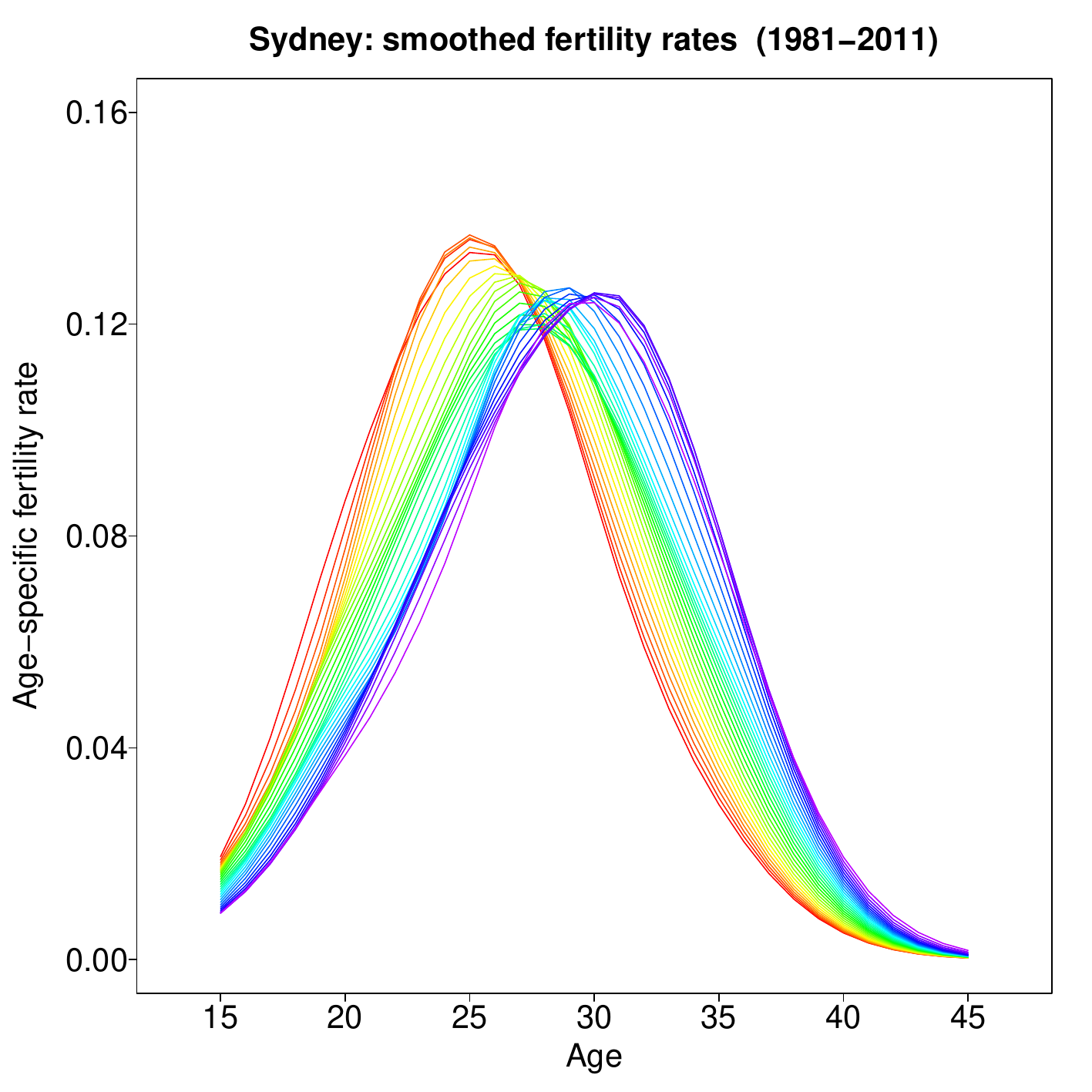}} 
\quad
\subfloat{\includegraphics[width = 3.5in]{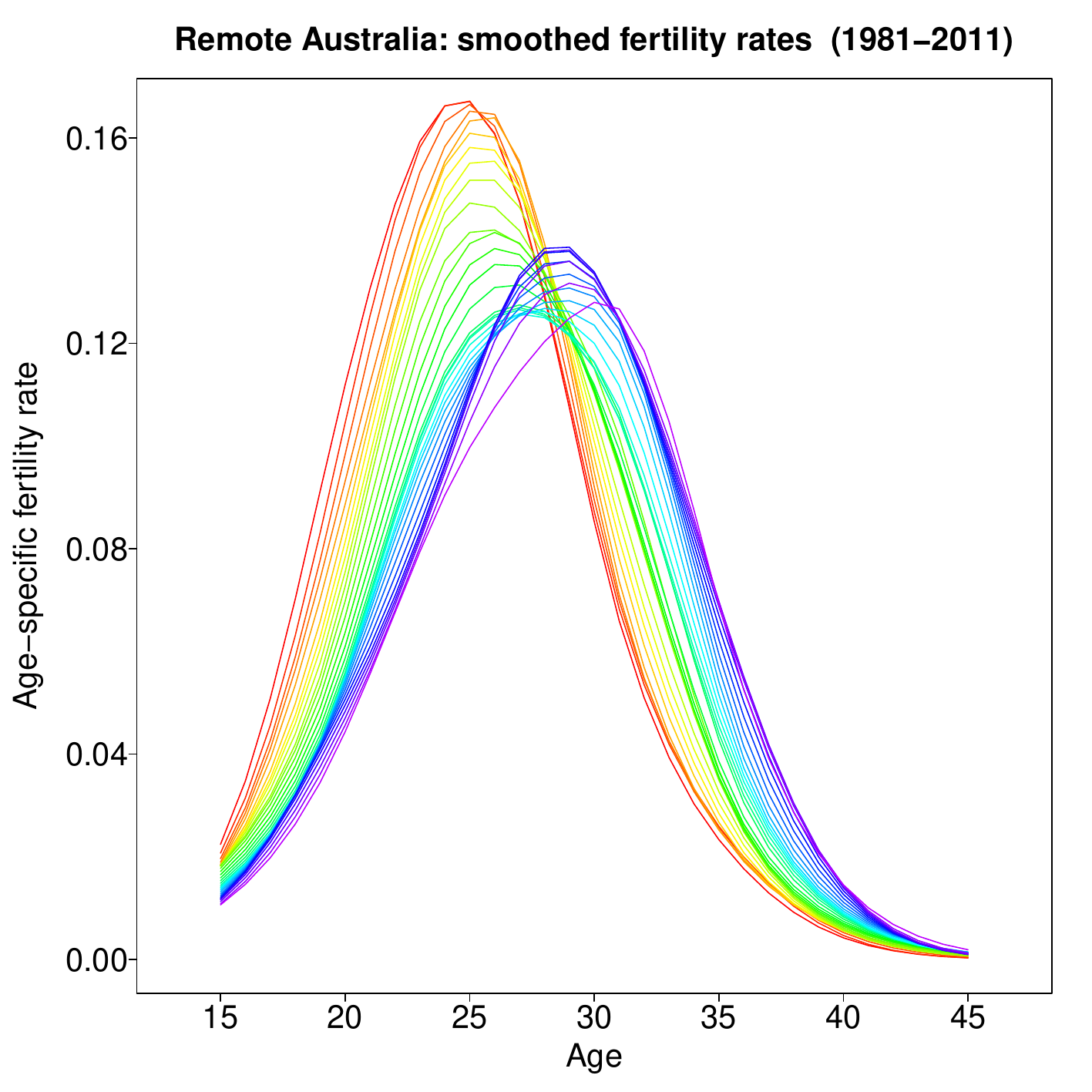}}
\caption{Selected age-specific fertility rates between 1981 and 2011 by administrative divisions} \label{fig:2}
\end{figure}

As a geographically large country, Australia has its populations in metropolitan cities and remote areas with quite different demographic features. For example, the Northern Territory has only 1\% of the country's population as of 2011. Still, a significantly higher percentage of the Aboriginal and Torres Strait Islander population, 28.1\% than the national average of 2.7\% \citep{ABSdemographic}. Figure~\ref{fig:2} compares smoothed age-specific fertility curves of remote Australia ($R_{11}$) to the region with the highest population in Australia, namely Sydney ($R_1$). 

Notably, remote Australia experienced a greater decline in peak age-specific fertility than Sydney between 1981 and 2000. This is because of high fertility among Indigenous women aged between 15--25 years, which is a well-documented pattern of the Aboriginal population, particularly living in the Northern Territory, continued to decline since the 1980s \citep{Smith1980, Johnstone2010}. Like remote Australia, other Australian regions and areas have their unique characteristics of age-specific fertility distribution, in addition to the common features of fertility development for all Australian women, illustrated in Figure~\ref{fig:1}. By considering administrative divisions at various levels, we aim to perceive spatial differences in making forecasts of the age-specific fertility for all sub-national populations in Australia.

\begin{figure}[!htb]
\centering
\subfloat{\includegraphics[width = 3.5in]{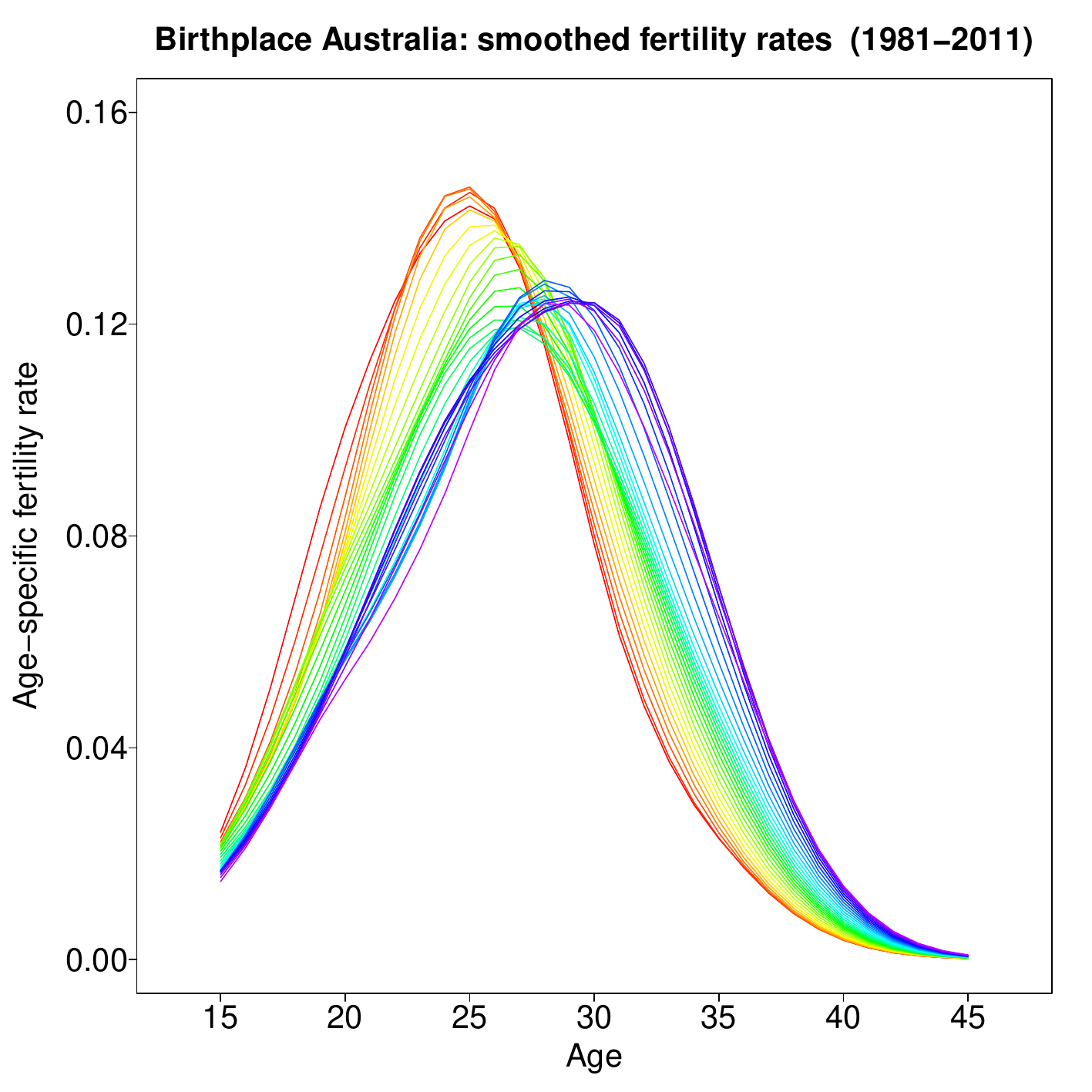}} 
\quad
\subfloat{\includegraphics[width = 3.5in]{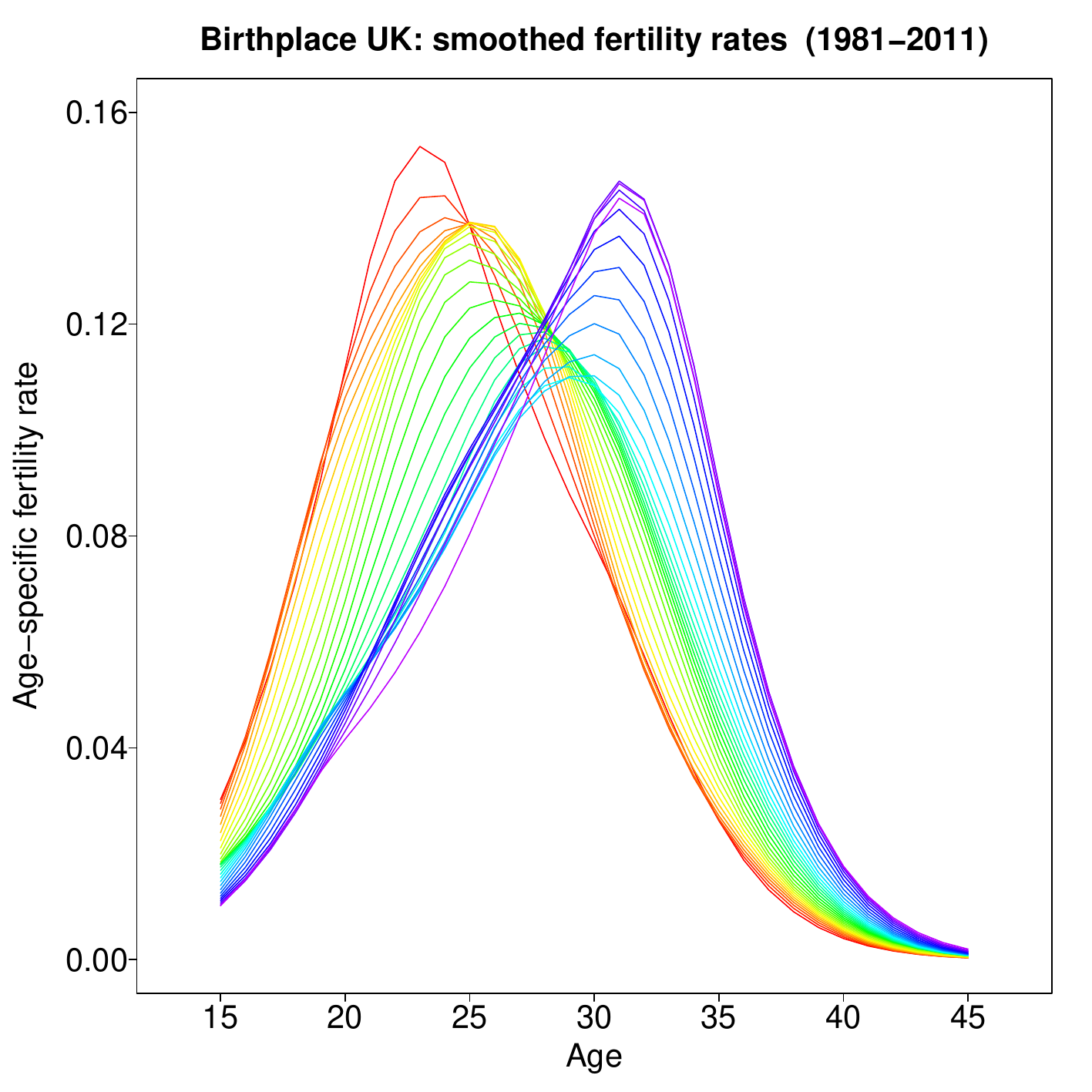}}
\caption{Selected age-specific fertility rates between 1981 and 2011 by birthplace}\label{fig:3}
\end{figure}

Australia is akin to most developed countries in the world, where immigrants add both labor and births to the population (see, e.g., \cite{MS12} in Italy; \cite{Andersson04} in Sweden; \cite{RB16} in England and Wales). For instance, in 1981, 68\% of immigration to Australia consisted of people born in Europe or New Zealand, and only 19\% were born in Asia; in 2016, the percentages were opposite, 21\% and 60\%, respectively \citep{BRV21}. Australia's total fertility rate has been below the replacement (around 2.1 babies per woman) required to replace herself and her partner and prevent a long-run population decline \citep{ABSbirth}. Australia's net overseas migration has generally been over $50,000$ since 1980 and exhibits a pattern of rapid growth during the first decade of the 21\textsuperscript{st} century \citep{ABSmigration}. Positive net overseas migration has been the most critical driver of Australia's population growth since the mid-2000s \citep{CP21planning}. By examining birthplace, we aim to understand which country of origin contributes to population growth in Australia through their births. Among the 18 different immigration groups in Australia, we found an apparent heterogeneity of immigrants and their contributions to the number of births in Australia from 1981 to 2011. Figure~\ref{fig:3} shows age-specific fertility curves of two places of birth for women covered by the census, namely Australia ($\text{COB}_1$) and the United Kingdom (the UK, $\text{COB}_4$). 

As a domestic birthplace, the COB of Australia covers approximately 73.9\% of female residents considered in this study. The COB of the UK is the largest foreign birthplace accounting for about 4.8\% of Australian women. Figure~\ref{fig:3} compares distributions of age-specific fertility curves of Australia-born women (left panel) with those of UK-born women (right panel). The peak fertility rates of Australian-born women have declined, and the peak ages of childbearing have increased over the period between 1981--2011. In contrast, the fertility curves of UK-born women in 1981 and 2011 have similar peak rates, despite an approximately 10-year difference in peak ages. Hence, the UK-born women in Australia appear to change the timing of births but not the total fertility rates. This postponement of childbearing of UK-born women reflected by the age-specific fertility curves is consistent with the increasing trend of most common age at childbirth in England and Wales \citep{ONS22}.

\section{Functional time series forecasting method} \label{sec:3}

Following \cite{SY19}, we use a hierarchical model coupled with grouped multivariate functional principal component decomposition technique to forecast age-specific fertility rates. The method models multiple sets of fertility functions correlated at any particular level of the chosen hierarchy. The common features in the data groups are considered in the estimation of the long-run covariance function, which is the sum of covariance function and autocovariance functions at various lags. The population-specific trend is captured by the principal components corresponding to the individual fertility series. Finally, forecasts of the empirical principal component scores are produced for each population before transforming back to the age-specific fertility rate forecasts. We present details of this functional time series forecasting method in the remaining of this section.

\begin{figure}[!htb]
\centering
\subfloat[Hierarchy according to administrative division]
{\includegraphics[width = 3.5in]{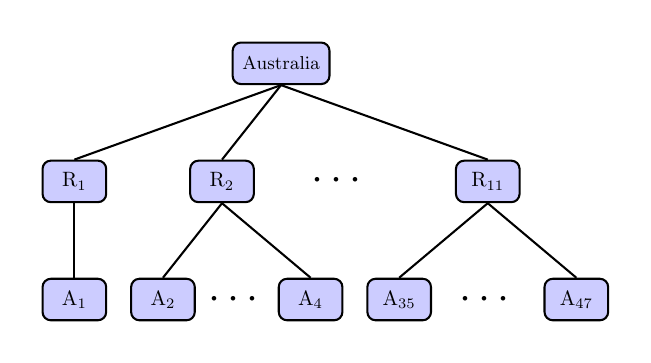}\label{fig:4a}} 
\quad
\subfloat[Hierarchy according to birthplace]
{\includegraphics[width = 3.5in]{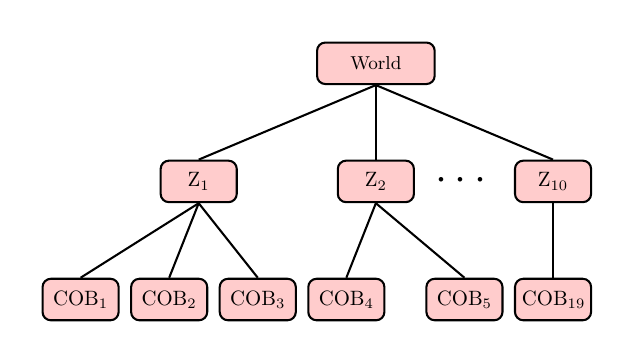}\label{fig:4b}}
\caption{Group structures for forecasting sub-national age-specific fertility rates of Australia via the grouped multivariate functional time series method}\label{fig:4}
\end{figure}

Using administrative division and birthplace as disaggregation factors, we can organize Australia's sub-national age-specific fertility rates into two group structures shown in Figure~\ref{fig:4}. Both structures comprise three levels, with parent nodes in the top level. The middle level consists of at least one child node. For example, in Figure~\ref{fig:4a}, the NSW Coast ($R_2$) region as a parent node is comprised of areas including Hunter ($A_2$), Illawarra ($A_3$), and Mid-North Coast ($A_4$) as children. When multiple children exist for any particular parent node, female populations represented by these children nodes are highly correlated spatially, either within Australia (administrative division hierarchy in Figure~\ref{fig:4a}) or around the globe (birthplace hierarchy in Figure~\ref{fig:4b}). Hence, fertility series of children nodes belonging to the same parent are collectively modeled via the grouped multivariate functional time series method \citep{CCY14, SY19, SK22}. We briefly describe this method to produce forecasts for Australian fertility rates.

Define  $\{ f^{(j)}(x) \}_{j = 1, \cdots, \omega }$ as a set of smoothed sub-national fertility functions, with $\omega$ representing the number of children nodes in consideration, and $\omega = 1$ corresponding to the special case of univariate functional time series. All $f^{(j)}(x)$ under consideration are square-integrable functions defined over the same interval of age $\mathcal{I} = [15,45]$. Let $\bm{f}(x) = \left[f^{(1)}(x),\dots,f^{(\omega)}(x)\right]^{\top}$ with $\omega \geq 2$ be a vector in the Hilbert space. Let $\mu^{(j)}(x) := \mathbb{E}[f^{(j)}(x)]$ denote the mean function for the $l^{\text{th}}$ population. The autocovariance matrix of $\bm{f}(x)$ at lag $\ell$ can be easily obtained as 
\begin{align*}
   \gamma_{\ell} (x,z) = \text{Cov} \left[\bm{f}_0(x), \bm{f}_{\ell}(z)\right],
\end{align*} 
with its $(l,j)^{\text{th}}$ ($1\leq l,j \leq \omega$) element given by
\begin{align*}
\gamma^{lj}_{\ell}(x, z) :=  \mathbb{E} \Big\{ [f_0^{(l)}(x)-\mu^{(l)}(x)] [f_{\ell}^{(j)}(z) - \mu^{(j)}(z)] \Big\}  = \text{Cov}\left[f_0^{(l)}(x), f_{\ell}^{(j)}(z)\right], \quad x, z\in \mathcal{I}.
\end{align*}
We can then obtain a long-run covariance function of $\bm{f}(x)$ as
\begin{align*}
  C_{lj}(x,z) = \sum_{\ell = -\infty}^{\infty} \gamma_{\ell} (x,z).
\end{align*}

For the $j^{\text{th}}$ population, denote a vector of long-run covariance functions as $\bm{C}_j(x,z)$ = $\Big[C_{j1}(x,z)$, $\cdots, C_{j \omega}(x,z) \Big]^{\top} $. We can then define an integral operator $\mathcal{A}: \mathbb{H} \rightarrow \mathbb{H} $ with the covariance kernel $\bm{C}(x,z) = \{C_{lj}(x,z), 1\leq l,j \leq \omega\}$ such that
 \begin{align*}
  (\mathcal{A} \bm{f} ) (x) = \int_{\mathcal{I}} \bm{C}(x,z) \bm{f} (z) dz = \begin{pmatrix}
    \inp{\bm{C}_1(x,z)}{\bm{f}(z)} \\ \vdots \\  \inp{\bm{C}_{\omega}(x,z)}{\bm{f}(z)}   
  \end{pmatrix},
 \end{align*}
where $\inp{\bm{C}_j(x,z)}{\bm{f}(z)} = \sum_{l=1}^{\omega}\inp{C_{jl}(x,z)}{f^{(l)}(z)} $ and $\bm{f} \in \mathbb{H}$. In practice, an empirical long-run covariance can be estimated as 
\begin{align*}
  \widehat{C}_{\mathcal{h}}(x,z) = \sum_{\ell = -n}^{n} W \left( \frac{\ell}{\mathcal{h}} \right) \widehat{\gamma}_{\ell}(x,z),
\end{align*}
where $W$ is symmetric weight function, $\mathcal{h}$ is a bandwidth parameter; and the estimator of $\gamma_{\ell}(x,z)$ is defined in the form of
\begin{equation*}
\widehat{\gamma}_{\ell}(x,z) = \begin{cases}
\frac{1}{n} \sum_{t=1}^{n-\ell} \left[\bm{f}_t(x) - \widehat{\bm{\mu}}(x) \right] \left[\bm{f}_{t+\ell}(z) - \widehat{\bm{\mu}}(z) \right], & \ell \geq 0; \\
\frac{1}{n} \sum_{t=1-\ell}^{n} \left[\bm{f}_t(x) - \widehat{\bm{\mu}}(x) \right] \left[\bm{f}_{t+\ell}(z) - \widehat{\bm{\mu}}(z) \right], & \ell < 0.
\end{cases}
\end{equation*}
The bandwidth parameter $\mathcal{h}$ in the kernel function can greatly affect the estimator's performance in finite samples. Therefore, we prefer to select $\mathcal{h}$ via a data-driven approach, such as the ``plug-in'' algorithm proposed in \cite{RS17}. 

By Mercer's lemma, there exists orthonormal sequences $\{\bm{\phi}_k = [\phi_k^{(1)}, \cdots, \phi_k^{(\omega)}]^{\top} \}_{k=1,2,\cdots}$ of continuous functions in $\mathbb{H}$, and a non-increasing sequence $\lambda_k$ of positive numbers, such that
 \begin{align*}
 \bm{C}(x,z) = \sum_{k=1}^{\infty} \lambda_k \bm{\phi}_k(x)\bm{\phi}_k(z),
 \end{align*}
 with the $(l,j)^{\text{th}}$ element $C_{lj}(x,z)$ of $\bm{C}(x,z) $
 \begin{align*}
  C_{lj}(x,z) = \sum_{k=1}^{\infty}\lambda_k \phi_{k}^{(l)}(x) \phi_{k}^{(j)}.
 \end{align*}
 By the Karhunen-Lo\`{e}ve theorem, the stochastic process for the $j^{\text{th}}$ population can be expressed as
 \begin{align}
 f^{(j)}(x) &= \mu^{(j)}(x) + \sum^{\infty}_{k=1}\beta^{(j)}_{k}\phi_k^{(j)}(x), \label{eq_2}
 \end{align}
 where $\beta^{(j)}_{k}$ is the $k^{\text{th}}$ principal component score given by the projection of $[f^{(j)}(x) - \mu^{(j)}(x)]$ in the direction of eigenfunction $\phi_k^{(j)}$, that is, $\beta^{(j)}_{k} = \inp{f^{(j)}(x) - \mu^{(j)}(x)}{\phi_k^{(j)}(x)}$.

Applying the decomposition of \eqref{eq_2} to a time series of smoothed fertility functions $\{\bm{f}_1(x), \cdots, \bm{f}_n(x)\}$ gives
\begin{align*} 
  f_t^{(j)}(x) &= \mu^{(j)}(x) + \sum^{\infty}_{k=1}\beta^{(j)}_{t,k}\phi_k^{(j)}(x)\\
  &= \mu^{(j)}(x) + \sum_{k=1}^{K}\beta^{(j)}_{t,k}\phi^{(j)}_k(x) + e^{(j)}_t(x),
 \end{align*}
where $\Big\{\phi_1^{(j)}(x),\dots,\phi_{K}^{(j)}(x)\Big\}$ is a set of orthogonal basis functions commonly known as functional principal components for the $l^{\text{th}}$ population, with $\Big\{\bm{\beta}_1^{(j)},\cdots,\bm{\beta}^{(j)}_{K}\Big\}$ their related principal component scores and $\bm{\beta}_k^{(j)} = \left[\beta_{1,k}^{(j)},\dots,\beta_{n,k}^{(j)}\right]^{\top}$ for $k = 1, \ldots, K$; $e_t^{(j)}(x)$ denotes the model truncation error function with mean zero and finite variance for the $j^{\text{th}}$ population. We select $K$ as the minimum of leading principal components reaching 95\% of total variance explained \citep{SH17}, such that 
\begin{equation*}
 K = \argmin_{K: K\geq 1}\left\{\sum^K_{k=1}\widehat{\lambda}_k\Bigg/\sum^{\infty}_{k=1}\widehat{\lambda}_k\mathds{1}_{\left\{\widehat{\lambda}_k>0\right\}}\geq 0.95\right\},
\end{equation*}
where $\mathds{1}\{\cdot\}$ represents the binary indicator function. There are alternative methods for determining the number of retained components, such as those of \cite{Chiou12, YMW05, RS91, HV06}, but they are beyond the main focus of this paper. 

Collectively modeling $\omega$ women populations requires truncating at the first $K^{\text{th}}$ functional principal components of all time series as
 \begin{equation*}
 \bm{f}_t(x) = \bm{\Phi}(x)\bm{\beta}_t,
 \end{equation*}
where $\bm{\beta}_t = \left\{ \beta_{t,1}^{(1)}, \dots, \beta_{t,K}^{(1)}, \beta_{t,1}^{(2)},\dots,\beta_{t,K}^{(2)},\dots, \beta_{t,1}^{(\omega)},\dots,\beta_{t,K}^{(\omega)}\right\}^{\top}$ is an $(\omega\times K) \times 1$ vector of principal component scores and 
 \begin{equation*}
 \bm{\Phi}(x) = \left( \begin{array}{ccccccccc}
 \phi_1^{(1)}(x) & \cdots & \phi_{K}^{(1)}(x) & 0 & \cdots & 0 & 0 & \cdots & 0 \\
 0 & \cdots & 0 & \phi_1^{(2)}(x) & \cdots & \phi_{K}^{(2)}(x) & 0 & \cdots & 0 \\
 \vdots & \vdots & \vdots & \vdots & \vdots & \vdots & \vdots & \vdots & \vdots \\
  0 & \cdots & 0 & 0 & \cdots & 0 & \phi_1^{(\omega)}(x) & \cdots & \phi_{K}^{(\omega)}(x)  \end{array} \right)
 \end{equation*}
a $\omega \times (\omega\times K)$ matrix of the associated basis functions. With the empirically estimated $\widehat{\bm{\Phi}}(x)$, we can then make $h$-step-ahead point forecasts as
 \begin{align*}
 \widehat{\bm{f}}_{n+h|n} = \mathbb{E}[\bm{f}_{n+h}(x)|\bm{f}_1(x),\cdots,\bm{f}_n(x);\widehat{\bm{\Phi}}(x)] = \widehat{\bm{\mu}}(x) + \widehat{\bm{\Phi}}(x) \widehat{\bm{\beta}}_{n+h|n,k},
 \end{align*}
where the empirical mean function is defined as $\widehat{\bm{\mu}}(x) = \frac{1}{n}\sum_{t=1}^{n}\bm{f}_t(x)$. We adopt a univariate time series forecasting method of \cite{HS09} to obtain the forecast principal component score $\widehat{\bm{\beta}}_{n+h|n,k}$ \citep[see, also][]{SH17, SY19}. The advantage of univariate forecasting method is its ability to handle nonstationary time series.

In addition to point forecasts, we also compute interval forecasts to assess forecast uncertainty. Specifically, the method of~\cite{ANH15} is considered to construct pointwise prediction intervals as follows.
\begin{enumerate}[1)]
\item Using all observed data, compute the empirical functional principal components $\widehat{\bm{\Phi}}(x)$ with their associated estimated principal component scores $\{\widehat{\bm{\beta}}_{1}, \cdots, \widehat{\bm{\beta}}_{n}\}$, where $\widehat{\bm{\beta}}_t = \big\{ \widehat{\bm{\beta}}_{t,1}^{(1)}, \dots, \widehat{\bm{\beta}}_{t,K}^{(1)},$ $\dots, \widehat{\bm{\beta}}_{t,1}^{(\omega)},\dots,\widehat{\bm{\beta}}_{t,K}^{(\omega)}\big\}^{\top}$.  The in-sample forecasts are then constructed as
\begin{equation*}
\widehat{\bm{f}}_{\xi+h}(x) = \widehat{\bm{\mu}}(x) + \widehat{\bm{\Phi}}(x) \widehat{\bm{\beta}}_{\xi+h,k}, \qquad \xi \in \{K, \cdots,n-h\},
\end{equation*}
where $\widehat{\bm{\beta}}_{\xi+h}$ are $h$-step-ahead forecasts produced by univariate time series models based on $\widehat{\bm{\beta}}_{\xi}$.
\item With the in-sample point forecasts, we calculate the in-sample point forecasting errors
\begin{equation*}
\widehat{\bm{\epsilon}}_{\zeta}(x) = \bm{f}_{\xi+h}(x) - \widehat{\bm{f}}_{\xi+h}(x),
\end{equation*}
where $\zeta \in \{1,2,\cdots, M \} $ and $M = n-h-K+1$.
\item Follow~\cite{SH18}, we also use the nonparametric bootstrap approach to calculate pointwise prediction intervals. Determine a $\pi_{\alpha}$ such that $\alpha \times 100\%$ of in-sample forecasting errors satisfy
\begin{equation*}
\pi_{\alpha} \times \gamma_{\text{lb}}(x_i) \leq \widehat{\bm{\epsilon}}_{\zeta}(x_i) \leq \pi_{\alpha} \times \gamma_{\text{ub}}(x_i), \qquad i = 1,\dots,p.
\end{equation*}
Then, the $h$-step-ahead pointwise prediction intervals are given as 
\begin{equation*}
\pi_{\alpha} \times \gamma_{\text{lb}}(x_i) \leq \bm{f}_{n+h}(x_i) - \widehat{\bm{f}}_{n+h|n}(x_i) \leq \pi_{\alpha} \times \gamma_{\text{ub}}(x_i),
\end{equation*}
\end{enumerate}
where $i$ symbolizes the discretized data points. 

The approaches described above can be applied to both grouped structures shown in Figure~\ref{fig:4} to compute point and interval forecasts at the most disaggregated levels. In this way, the base forecasts may not add to the national-level forecasts. In Section~\ref{sec:4}, we will discuss forecast reconciliation methods used to address this issue.

\section{Forecast reconciliation methods}\label{sec:4}

We consider administrative division and birthplace as disaggregation factors for modeling sub-national fertility rates in Australia. For a particular $j$\textsuperscript{th} ($j \in \left\{\text{COB}_1, \ldots, \text{COB}_{19}, Z_1, \ldots, Z_{10}, \text{World}\right\} $) birthplace, denote all the base forecasts at the Area level in year $t$ as $\bm{A}_{j,t} = \left\{f_{\text{A1}\ast \text{j},t}, \ldots, f_{\text{A47}\ast \text{j},t} \right\}^{\top}$. Following the grouping structure shown in Figure~\ref{fig:4a}, we can express $\bm{f}_{j,t}$, a vector containing all series at all administrative division levels, as a product of matrices given by
\arraycolsep=0.04cm
\[
\underbrace{ \left[
\begin{array}{c}
f_{\text{AUS}\ast \text{j},t} \\
f_{\textcolor{a0}{\text{R1}\ast \text{j},t}} \\
f_{\textcolor{a0}{\text{R2}\ast \text{j},t}} \\
\vdots \\
f_{\textcolor{a0}{\text{R11}\ast \text{j},t}} \\
f_{\textcolor{blue}{\text{A1}\ast \text{j},t}} \\
f_{\textcolor{blue}{\text{A2}\ast \text{j},t}} \\
\vdots \\
f_{\textcolor{blue}{\text{A47}\ast \text{j},t}} \\ 
\end{array}
\right]}_{\bm{f}_{j,t}} =
\underbrace{\left[
\begin{array}{cccccccc}
\frac{\E_{\text{A1}\ast \text{j},t}}{\E_{\text{AUS}\ast \text{j},t}} & \frac{\E_{\text{A2}\ast \text{j},t}}{\E_{\text{AUS}\ast \text{j},t}} & \frac{\E_{\text{A3}\ast \text{j},t}}{\E_{\text{AUS}\ast \text{j},t}}  & \frac{\E_{\text{A4}\ast \text{j},t}}{\E_{\text{AUS}\ast \text{j},t}} & \cdots & \frac{\E_{\text{A47}\ast \text{j},t}}{\E_{\text{AUS}\ast \text{j},t}} \\
\textcolor{a0}{\frac{\E_{\text{A1}\ast \text{j},t}}{\E_{\text{R1}\ast \text{j},t}}} & \textcolor{a0}{0} & \textcolor{a0}{0} & \textcolor{a0}{0} & \cdots & \textcolor{a0}{0} \\
\textcolor{a0}{0} & \textcolor{a0}{\frac{\E_{\text{A2}\ast \text{j},t}}{\E_{\text{R2} \ast \text{j},t}}} & \textcolor{a0}{\frac{\E_{\text{A3}\ast \text{j},t}}{\E_{\text{R2} \ast \text{j},t}}} & \textcolor{a0}{\frac{\E_{\text{A4}\ast \text{j},t}}{\E_{\text{R2}\ast \text{j},t}}} & \cdots & \textcolor{a0}{0} \\
\vdots & \vdots & \vdots & \vdots & \vdots & \vdots \\
\textcolor{a0}{0} & \textcolor{a0}{0} & \textcolor{a0}{0} & \textcolor{a0}{0} & \cdots & \textcolor{a0}{\frac{\E_{\text{A47}\ast \text{j},t}}{\E_{\text{R11}\ast \text{j},t}}} \\
\textcolor{blue}{1} & \textcolor{blue}{0} & \textcolor{blue}{0} & \textcolor{blue}{0} & \cdots & \textcolor{blue}{0} \\
\textcolor{blue}{0} & \textcolor{blue}{1} & \textcolor{blue}{0} & \textcolor{blue}{0} & \cdots & \textcolor{blue}{0} \\
\vdots & \vdots & \vdots & \vdots & \vdots & \vdots \\
\textcolor{blue}{0} & \textcolor{blue}{0} & \textcolor{blue}{0} & \textcolor{blue}{0} & \cdots & \textcolor{blue}{1} \\
\end{array}
\right]}_{\bm{S}_{j,t}}
\underbrace{\left[
\begin{array}{c}
f_{\text{A1}\ast \text{j},t} \\
f_{\text{A2}\ast \text{j},t} \\
\vdots \\
f_{\text{A47}\ast \text{j},t} \\
\end{array}
\right]}_{\bm{A}_{j,t}}
\]
\hspace{-.05in}
or $\bm{f}_{j,t} = \bm{S}_{j,t}\bm{A}_{j,t}$, where $\bm{S}_{j,t}$ is a summing matrix whose elements include ratios of population at risk in various administrative divisions. A general notation for population at risk is given by $\E_{\text{g} \ast j, t}$, where $g$ stands for a particular administrative division listed in Table~\ref{tab:2}. For example, $\E_{\text{A1}\ast \text{j},t}$ is the number of women in Sydney ($\text{Area}_1$) born in the $j$\textsuperscript{th} birthplace in year $t$, while $\E_{\text{AUS}\ast \text{j},t}$ represents the overall population of female residents in Australia born in the same place. 

The grouping structure shown in Figure~\ref{fig:4b} splits Australian fertility rates at the national level by birthplaces. For a particular $g$\textsuperscript{th} ($g \in \left\{A_1, \ldots, A_{47}, R_1, \ldots R_{11}, \text{Australia}\right\}$) administrative division, denote all the base forecasts at the COB level in year $t$ as $\bm{\text{COB}}_t = \left\{f_{\text{g}\ast \text{COB1},t}, \ldots, f_{\text{g}\ast \text{COB19},t} \right\}^{\top}$. We can express $\bm{f}_{g,t}$, a vector containing all series at all birthplaces, into a product of matrices as
\arraycolsep=0.04cm
\[
\underbrace{ \left[
\begin{array}{c}
f_{\text{g}\ast \text{World},t} \\
f_{\textcolor{a0}{\text{g}\ast \text{Z1},t}} \\
f_{\textcolor{a0}{\text{g}\ast \text{Z2},t}} \\
\vdots \\
f_{\textcolor{a0}{\text{g}\ast \text{Z{10}},t}} \\
f_{\textcolor{blue}{\text{g}\ast \text{COB1},t}} \\
f_{\textcolor{blue}{\text{g}\ast \text{COB2},t}} \\
\vdots \\
f_{\textcolor{blue}{\text{g}\ast \text{COB19},t}} \\ 
\end{array}
\right]}_{\bm{f}_{g,t}} =
\underbrace{\left[
\begin{array}{cccccccc}
\frac{\E_{\text{g}\ast \text{COB1},t}}{\E_{\text{g}\ast \text{World},t}} & \frac{\E_{\text{g}\ast \text{COB2},t}}{\E_{\text{g}\ast \text{World},t}} & \frac{\E_{\text{g}\ast \text{COB3},t}}{\E_{\text{g}\ast \text{World},t}}  & \frac{\E_{\text{g}\ast \text{COB4},t}}{\E_{\text{g}\ast \text{World},t}} & \cdots & \frac{\E_{\text{g}\ast \text{COB19},t}}{\E_{\text{g}\ast \text{World},t}} \\
\textcolor{a0}{\frac{\E_{\text{g}\ast \text{COB1},t}}{\E_{\text{g}\ast \text{Z1},t}}} & \textcolor{a0}{\frac{\E_{\text{g}\ast \text{COB2},t}}{\E_{\text{g}\ast \text{Z1},t}}} & \textcolor{a0}{\frac{\E_{\text{g}\ast \text{COB3},t}}{\E_{\text{g}\ast \text{Z1},t}}} & \textcolor{a0}{0} & \cdots & \textcolor{a0}{0} \\
\textcolor{a0}{0} & \textcolor{a0}{0} & \textcolor{a0}{0} & \textcolor{a0}{\frac{\E_{\text{g}\ast \text{COB4},t}}{\E_{\text{g}\ast \text{Z2},t}}} & \cdots & \textcolor{a0}{0} \\
\vdots & \vdots & \vdots & \vdots & \vdots & \vdots \\
\textcolor{a0}{0} & \textcolor{a0}{0} & \textcolor{a0}{0} & \textcolor{a0}{0} & \cdots & \textcolor{a0}{\frac{\E_{\text{g}\ast \text{COB19},t}}{\E_{\text{g}\ast \text{Z10},t}}} \\
\textcolor{blue}{1} & \textcolor{blue}{0} & \textcolor{blue}{0} & \textcolor{blue}{0} & \cdots & \textcolor{blue}{0} \\
\textcolor{blue}{0} & \textcolor{blue}{1} & \textcolor{blue}{0} & \textcolor{blue}{0} & \cdots & \textcolor{blue}{0} \\
\vdots & \vdots & \vdots & \vdots & \vdots & \vdots \\
\textcolor{blue}{0} & \textcolor{blue}{0} & \textcolor{blue}{0} & \textcolor{blue}{0} & \cdots & \textcolor{blue}{1} \\
\end{array}
\right]}_{\bm{S}_{g,t}}
\underbrace{\left[
\begin{array}{c}
f_{\text{g}\ast \text{COB1},t} \\
f_{\text{g}\ast \text{COB2},t} \\
\vdots \\
f_{\text{g}\ast \text{COB19},t} \\
\end{array}
\right]}_{\bm{\text{COB}}_{g,t}}
\]
\hspace{-.05in} 
or $\bm{f}_{g,t} = \bm{S}_{g,t} \bm{\text{COB}}_{g,t}$. The summing matrix $\bm{S}_{g,t}$ is designed in the same fashion as $\bm{S}_{j,t}$, except for fixing a particular administrative admission $g$.

To ease notations, without causing confusion we may use general notations $\bm{B}_t$ for base level forecasts ($\bm{A}_{j,t}$ or $\bm{\text{COB}}_{g,t}$), $\bm{S}_t$ for either summing matrices ($\bm{S}_{j,t}$ or $\bm{S}_{g,t}$), and $\bm{f}_t$ for the reconciled forecasts at all levels of disaggregation ($\bm{f}_{j,t}$ or $\bm{f}_{g,t}$), respectively. In Sections~\ref{sec:4.1} and~\ref{sec:4.2}, we shall present two forecasts reconciliation methods using these general notations.

\subsection{Bottom-up method} \label{sec:4.1}

The bottom-up (BU) method has long been used to reconcile forecasts in hierarchical structures \citep[see, e.g.,][]{DM92, ZT00,SH17}. Specifically, the method aggregates base forecasts at the most disaggregated level upward towards the national total. For instance, let us consider the Australian fertility data divided by administrative divisions illustrated in Figure~\ref{fig:4a}. Focusing on fertility series of a particular birthplace $j$, we first generate $h$-step-ahead base forecasts for the most disaggregated series, namely $\widehat{\bm{B}}_{n+h|n} = \left\{\widehat{f}_{A1 \ast j, n+h|n},\dots, \widehat{f}_{A47 \ast j, n+h|n}\right\}^{\top}$. We then proceed to obtain reconciled forecasts $\overline{\bm{f}}_{n+h|n}$ for all series as
\begin{align*}
  \overline{\bm{f}}_{n+h|n} = \widehat{\bm{S}}_{n+h|n}\widehat{\bm{B}}_{n+h|n},
\end{align*}
where $\widehat{\bm{S}}_{n+h|n}$ denotes the $h$-step-ahead of the summing matrices. Based on historical ratios of population at risk, elements in $\widehat{\bm{S}}_{n+h|n}$ are forecast by an automated autoregressive integrated moving average algorithm of \cite{Hyndman2008}. Specifically, let $p_t = \E_{X\ast W,t}/\E_{Y\ast W',t} $ be a nonzero ratio in $\bm{S}_t$. Given the observed time series $\{p_1, \ldots, p_n\}$, we make $h$-step-ahead forecast to obtain $\widehat{p}_{n+h|n}$. Repeat the process for each ratio in the summing matrix yields the $\widehat{\bm{S}}_{n+h|n}$ used in $h$-step-ahead forecasts reconciliation. This method of obtaining forecast summing matrices is commonly used in grouped functional time series forecasting applications \citep[see, e.g.,][]{SH17}.

\subsection{Optimal combination method} \label{sec:4.2}

Instead of considering only the bottom-level series, \cite{HAA+11} put forward a method where base forecasts for all series are computed before reconciling the forecasts so that they satisfy the aggregation constraints. The optimal combination method assembles the base forecasts through a linear regression, where the reconciled forecasts are close to the base forecasts while aggregating consistently. 

Stacking the $h$-step-ahead base forecasts for all series in the same order as for the original data and denote the resulted vector as $\widehat{\bm{f}}_{n+h|n}$. The optimal combination method considers these base forecasts as the response variable in a linear regression given by
\begin{align*}
  \widehat{\bm{f}}_{n+h|n} = \widehat{\bm{S}}_{n+h|n}\bm{\beta}_{n+h} + \bm{\varepsilon}_{n+h},
\end{align*}
where $\bm{\beta}_{n+h} = \mathbb{E}\left[\bm{B}_{n+h|n}|\bm{f}_1, \ldots, \bm{f}_n\right]$ is the unknown mean of the forest distributions of the most disaggregated series, and $\bm{\varepsilon}_{n+h}$ is the reconciliation error with mean zero and variance-covariance matrix $\Sigma_h = \text{var}(\bm{\varepsilon}_{n+h})$. Following \cite{HLW16}, we estimate the regression coefficient $\bm{\beta}_{n+h}$ via least squares as
\begin{align*}
  \widehat{\bm{\beta}}_{n+h|n} = \left(\widehat{\bm{S}}_{n+h|n}^{\top}\bm{W}^{-1}\widehat{\bm{S}}_{n+h|n}\right)^{-1}\widehat{\bm{S}}_{n+h|n}^{\top}\bm{W}^{-1}\widehat{\bm{R}}_{n+h|n},
\end{align*}
where $\bm{W}$ is a weight matrix customarily selected as $\bm{W} = c \times \bm{I}$ with any constant $c$ and the identity matrix $\bm{I}$. Finally, we compute the reconciled forecasts as
\begin{align*}
  \overline{\bm{f}}_{n+h|n} = \widehat{\bm{S}}_{n+h|n}\widehat{\bm{\beta}}_{n+h} = \widehat{\bm{S}}_{n+h|n}\left(\widehat{\bm{S}}_{n+h|n}^{\top}\widehat{\bm{S}}_{n+h|n}\right)^{-1}\widehat{\bm{S}}_{n+h|n}^{\top}\widehat{\bm{B}}_{n+h|n}.
\end{align*}

\subsection{Trace minimization method} \label{sec:4.3}

A more recent trace minimization (MinT) method proposed by \cite{WAH19} aims at minimizing the sum of variances of reconciliation errors to obtain coherent forecasts across the entire collection of time series. The summing matrix $\widehat{\bm{S}}_{n+h|n}$ in our application varies with time. For this reason, we extend the MinT method to adapt to the time-varying covariance of $h$-step-ahead forecast errors. 

Define $\I_t = \left\{ \bm{f}_{1}, \ldots, \bm{f}_t; \bm{E}_{1}, \ldots, \bm{E}_t \right\}$ where $\bm{E}_t$ represents ERP exposures in year $t$ at all levels of the considered hierarchy. First, compute errors of the $h$-step-ahead reconciled forecasts as
\begin{align*}
  \widetilde{\bm{e}}_{t+h} = \bm{f}_{t+h} - \overline{\bm{f}}_{t+h|t}.
\end{align*}

We assume that the time series of summing matrices $\{\bm{S}_t: t \in \mathbb{Z}\}$ is stationary with  $\mathbb{E}\left[\bm{S}_{t+h} | \I_t  \right] = \mathbb{E}\left[\bm{S}_{t}\right] = \widetilde{\bm{S}} $ for a small $h \in \mathbb{Z}^{+}$. It can then be shown that
\begin{align*}
  \text{var}\left[\widetilde{\bm{e}}_{t+h} | \I_t \right] \approx \widetilde{\bm{S}} \bm{P}_{t+h} \bm{W}_h \bm{P}_{t+h}^{\top} \widetilde{\bm{S}}^{\top},
\end{align*}
where $\bm{P}_{t+h}$ is a fixed projection matrix, and $\bm{W}_h = \mathbb{E}[\widehat{\bm{e}}_{t+h|t} \widehat{\bm{e}}_{t+h|t}^{\top}|\I_t]$ with $h$-step-ahead forecast errors $\widehat{\bm{e}}_{t+h|t} = \bm{f}_{t+h} - \widehat{\bm{f}}_{t+h|t}$. The projection matrix further satisfies that $\bm{P}_{t+h} \widetilde{\bm{S}} = \bm{I}$. For a particular $t$ and $h$, minimizing the trace of $\text{var}\left[\widetilde{\bm{e}}_{t+h}|\I_t\right]$ yields the projection matrix
\begin{align}
  \bm{P}_{t+h} = \left(\widetilde{\bm{S}}^{\top}{\bm{W}}_h^{-1}\widetilde{\bm{S}}\right)^{-1}\widetilde{\bm{S}}^{\top}{\bm{W}}_h^{-1}. \label{eq_3}
\end{align}
More details on the derivation of \eqref{eq_3} are provided in the Appendix.

When an ARIMA model paired with the mean squared error loss function are used to forecast the $h$-step-ahead summing matrix, the optimal forecast obtained is the conditional mean, i.e., $\widehat{\bm{S}}_{n+h|n} = \mathbb{E}\left[\bm{S}_{n+h} | \I_n \right] = \widetilde{\bm{S}}$. Therefore, the MinT method produces the reconciled forecasts as
 \begin{align*}
  \overline{\bm{f}}_{n+h|n}  = \widehat{\bm{S}}_{n+h|n}\left(\widehat{\bm{S}}_{n+h|n}^{\top}{\bm{W}}_h^{-1}\widehat{\bm{S}}_{n+h|n}\right)^{-1}\widehat{\bm{S}}_{n+h|n}^{\top}{\bm{W}}_h^{-1} \widehat{\bm{f}}_{n+h|n}.
 \end{align*}
The covariance matrix of base forecasts errors, namely $\bm{W}_h$, is practically estimated by shirking off-diagonal entries of the unbiased covariance estimator of the in-sample one-step-ahead base forecast errors toward targets on its diagonal, with alternative estimations discussed in \citet[][Section 2.4]{WAH19}. 

\section{Empirical data analysis results}\label{sec:5}

The grouped functional time series forecasting method is applied to Australian sub-national fertility rates following both structures, illustrated in Figure~\ref{fig:4}. We then reconcile the obtained base forecasts via the bottom-up and optimal combination methods. To assess model and parameter stabilities over time, we consider an expanding window analysis of considered time series models \citep[see][Chapter 9 for details]{ZW06}, which is carried out in the following steps:
\begin{enumerate}
\item[1)] Determine a particular grouped structure to disaggregate Australian total fertility rates. Specifically, birthplace is considered fixed when the group structure in Figure~\ref{fig:4a} is used, whereas administrative division is deemed to be fixed when the group structure in Figure~\ref{fig:4b} is used.
\item[2)] Using fertility rates in 1981--2001 as a training set, we initially produce one- to 10-step ahead point and interval forecasts before reconciliation via the base, bottom-up, and optimal combination methods.
\item[3)] Increase the training set with one more year of observations (i.e., fertility rates in 1981--2002), produce one-to nine-step-ahead point and interval forecasts, and compute the reconciled forecasts.
\item[4)] Iterate the process with the size of the training set increased by one year each time until reaching the end of the data period in 2011.
\end{enumerate}
The process produces 10 one-step-ahead forecasts, 9 two-step-ahead forecasts, and so on, up to one 10-step-ahead forecast. We report evaluation results of point forecast accuracy and interval forecast accuracy in Sections~\ref{sec:5.1} and~\ref{sec:5.2}, respectively.

\subsection{Comparison of point forecast accuracy}\label{sec:5.1}

We evaluate the accuracy of point forecasts by mean absolute scaled error (MASE), which measures the closeness of the forecasts compared to the actual values of the variable being forecast. Since each of the raw fertility series contains seven age groups (i.e., 15--19, 20--24, $\ldots$, 45--49) for any year between 2002 and 2011, we only compute MASE at these grid points as 
\begin{align*}
\text{MASE}^{(j)}(h) &= \frac{1}{7\times (11-h)}\sum^{10}_{\varsigma = h}\sum^{7}_{i=1} \frac{\abs{f^{(j)}_{n+\varsigma}(x_i) - \widehat{f}^{(j)}_{n+\varsigma}(x_i)} \mathds{1}_{\left\{f^{(j)}_{n+\varsigma}(x_i)>0\right\}}}{\frac{1}{n+\varsigma-h-1}\sum_{t=2}^{n+\varsigma-h-1}\abs{f_t^{(j)}(x_i)-f_{t-1}^{(j)}(x_i)}},
\end{align*}
where $f_{t}^{(j)}(x_i)$ represents the actual holdout sample for the $j$\textsuperscript{th} subpopulation in year $t$, and $\widehat{f}_t(x_i)$ represents the corresponding point forecasts. 

\begin{figure}[!htb]
\centering
\subfloat{\includegraphics[width = 7.3in]{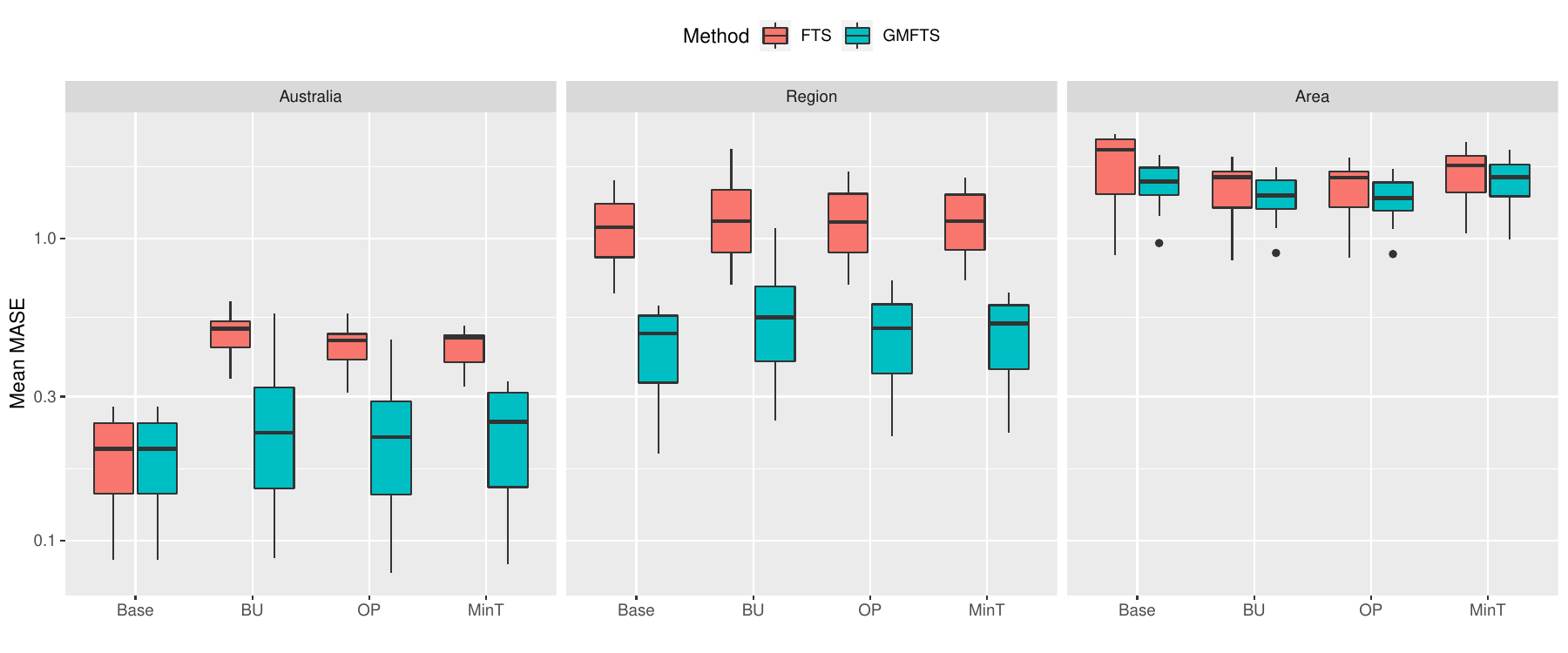}} \\
\subfloat{\includegraphics[width = 7.3in]{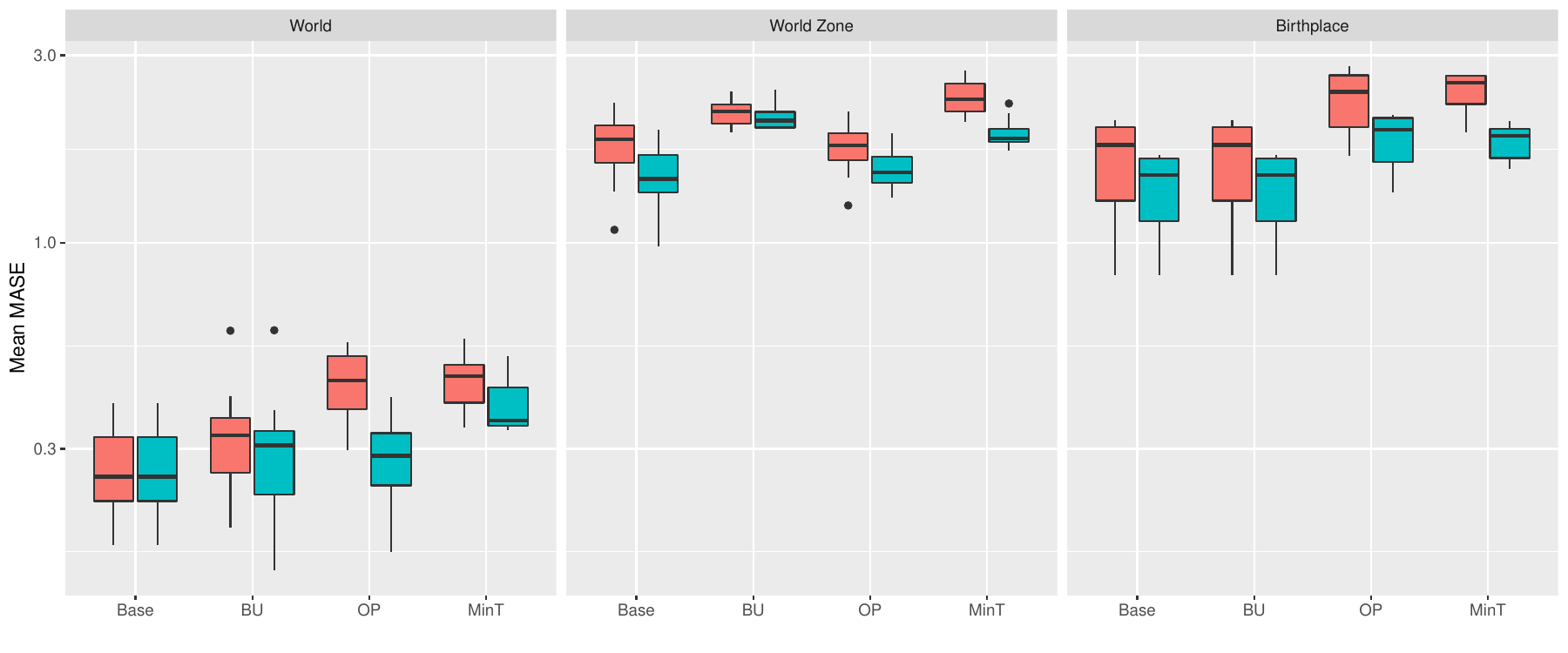}} 
\caption{Point forecast accuracy between the independent functional time series forecasting method labeled as FTS, and grouped functional time series forecasting methods labeled as GMTS} \label{fig:5}
\end{figure}

At a particular disaggregation level, averaging measurement of point forecasts over the total $\omega$ series (e.g., the area level in Figure~\ref{fig:4a} has $\omega = 47$) and 10 forecast horizons leads to a mean MASE given by
\begin{align*}
  \text{Mean(MASE)} = \frac{1}{10\omega} \sum_{h=1}^{10} \sum_{j=1}^{\omega} \text{MASE}^{(j)}(h).
\end{align*}
To facilitate a comparison of point forecasts at national and sub-national levels, we present the mean MASE of the base and reconciled forecasts related to the administrative division and birthplace group structures in the top panel and the bottom panel of Figure~\ref{fig:5}, respectively. Collectively modeling sub-national fertility rates lead to more accurate point forecasts following either hierarchy. This is because of significant heterogeneity associated with fertility series at the most disaggregated levels (e.g., the area or birthplace level). Pooling information of closely related subpopulations reduces the influences of abrupt and temporal movements of observations. It smooths out a part of the observational noise, subsequently improving the extraction of common features in Australian fertility. For example, we can measure the variability of fertility series related to all seven age groups in the area $j$ between 1981 and 2011 by the averaged median absolute deviation (MAD) as 
\begin{align*}
   \overline{\text{MAD}}_{j} = \frac{1.4826}{7}\sum_{i=1}^{7} \text{median} \left(\left|f^{(j)}_t(x_i) - \frac{1}{31}\sum_{t=1981}^{2011}f^{(j)}_t(x_i)\right|\right),
 \end{align*} 
where 1.4826 is a constant scale factor, suited for normally distributed data. Following this criterion, fertility time series of South West Queensland ($\text{Area}_{22}$) has the highest variability with $\overline{\text{MAD}}_{22} = 0.0163$, which is about 30\% more volatile than that of Sydney ($\text{Area}_1$). 

Comparing the mean MASE related to the two disaggregation group structures, the administrative division group structure generally leads to more accurate reconciled point forecasts than the birthplace group structure. This is due to some World Zones and birthplaces, e.g., Sub-Saharan Africa ($Z_{10}$ and $\text{COB}_{19}$), having a tiny female population with relatively large variability in the number of newborns over the census period. In contrast, all areas, including those in remote Australia, gradually increased with time and slowly changed fertility rates between 1981 and 2011. Thus, we can extract common features of fertility curves from all areas than from all birthplaces, resulting in more accurate point forecasts.

\subsection{Comparison of interval forecast accuracy}\label{sec:5.2}

To evaluate interval forecast accuracy, we utilize the interval score of \cite{GR07}. For each year in the forecasting period, the pointwise prediction intervals were computed at the $(1-\alpha)\times 100\%$ nominal coverage probability. Let $\widehat{f}_{n+h}^{(j),\text{lb}}(x)$ and $\widehat{f}_{n+h}^{(j),\text{ub}}(x)$ be the lower and upper bounds, respectively. A scoring rule for the interval forecast at discretized point $x_i$ is
\begin{align*}
S_{\alpha}\left[\widehat{f}_{n+h}^{(j),\text{lb}}(x_i), \widehat{f}_{n+h}^{(j),\text{ub}}(x_i), f_{n+h}^{(j)}(x_i)\right] = & \left[\widehat{f}_{n+h}^{(j),\text{ub}}(x_i) - \widehat{f}_{n+h}^{(j),\text{lb}}(x_i)\right] \\
& + \frac{2}{\alpha}\left[\widehat{f}^{(j),\text{lb}}_{n+h}(x_i) - f_{n+h}^{(j)}(x_i)\right]\mathds{1}\left\{f_{n+h}^{(j)}(x_i)<\widehat{f}_{n+h}^{(j),\text{lb}}(x_i)\right\} \\
& + \frac{2}{\alpha}\left[f_{n+h}^{(j)}(x_i) - \widehat{f}_{n+h}^{(j),\text{ub}}(x_i)\right]\mathds{1}\left\{f_{n+h}^{(j)}(x_i)>\widehat{f}_{n+h}^{(j),\text{ub}}(x_i)\right\},
\end{align*}
where $\mathds{1}\{\cdot\}$ represents the binary indicator function, and $\alpha$ denotes a level of significance. 

In this study, we utilize the nonparametric bootstrap approach to obtain in total $G=1000$ pointwise forecasts $f_{n+h}^{(j),g}(x_i)$ for any considered $j,h,i$. In the $g$\textsuperscript{th} ($g = 1,2,\ldots, 1000$) iteration, we firstly reconcile forecasts at all sub-national levels with a selected reconciliation method described in Section~\ref{sec:4}, and obtain forecasts $\widetilde{f}_{n+h}^{(j),g}(x_i)$. Next, averaging over all 10 years in the forecasting period gives the mean interval score for the total $\omega$ series as
\begin{align*}
  \overline{S}_{\alpha} = \frac{1}{10p\omega}\sum^{10}_{h=1} \sum_{i=1}^{p} \sum_{j=1}^{\omega} S_{\alpha}\left[\widehat{f}_{n+h}^{(j),\text{lb}}(x_i), \widehat{f}_{n+h}^{(j),\text{ub}}(x_i), \widetilde{f}_{n+h}^{(j),g}(x_i)\right].
\end{align*}

\begin{figure}[!htb]
\centering
\subfloat{\includegraphics[width = 7.3in]{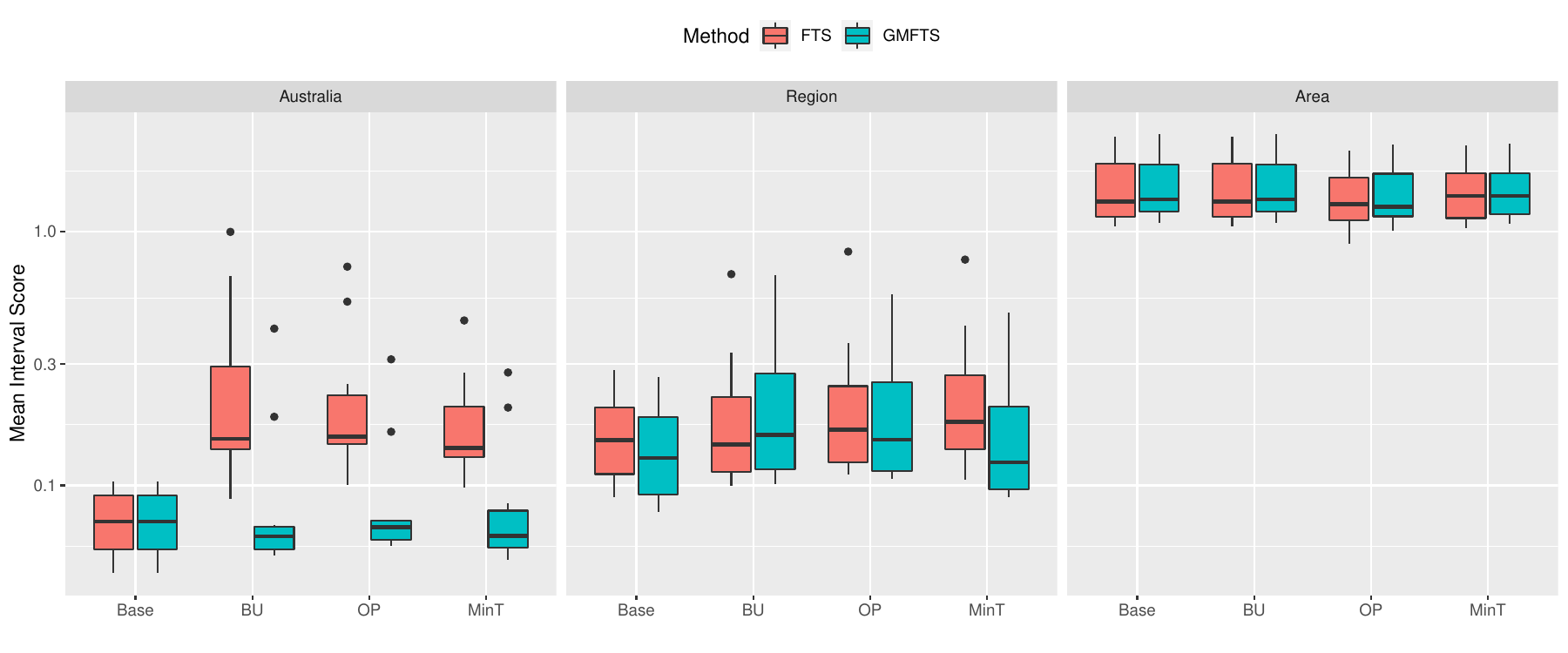}} \\
\subfloat{\includegraphics[width = 7.3in]{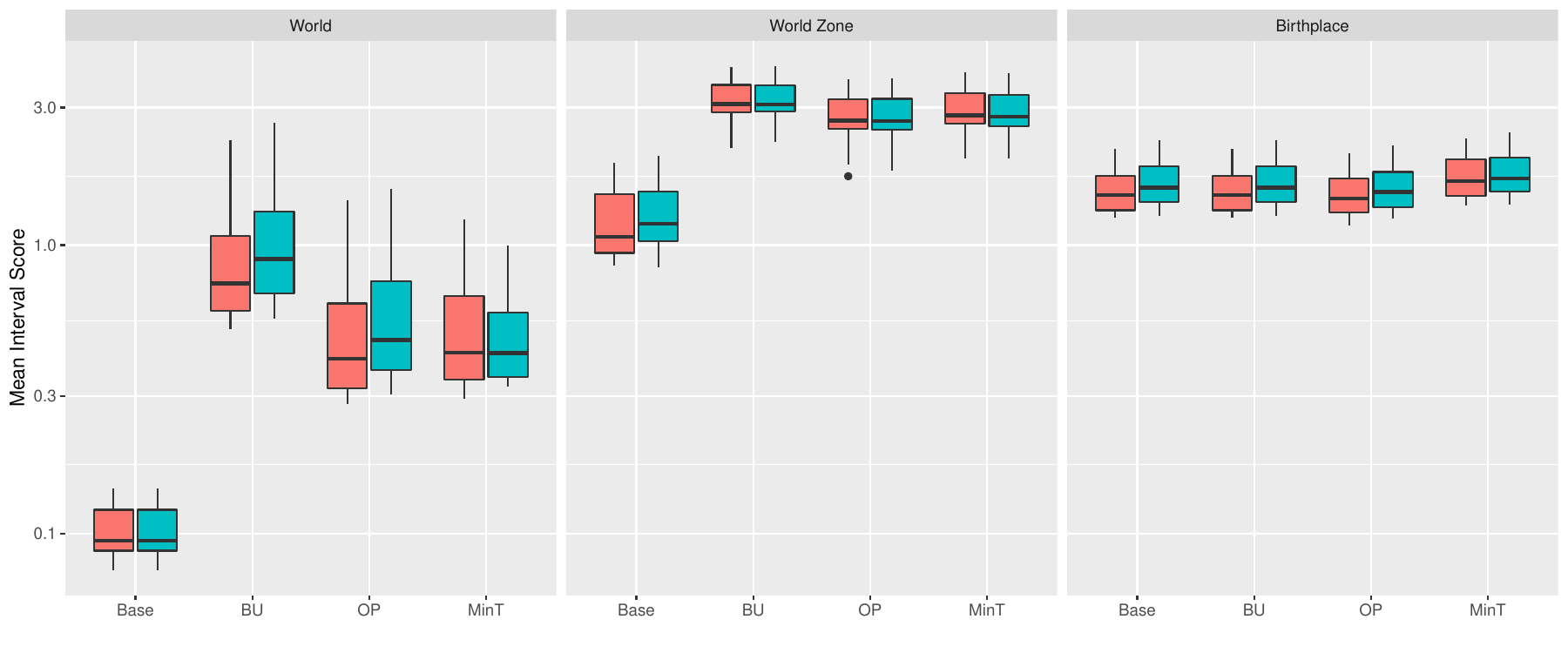}} 
\caption{Interval forecast accuracy between the independent and grouped functional time series forecasting methods} \label{fig:6}
\end{figure}

We present mean interval scores of base and reconciled forecasts related to the administrative division and birthplace group structures in the top and bottom panels of Figure~\ref{fig:6}, respectively. Comparing the mean interval scores related to the two disaggregation group structures, the administrative division group structure generally leads to more accurate reconciled interval forecasts than the birthplace group structure. In contrast to the point forecast results, the reconciled interval forecasts do not always improve forecast accuracy compared to the base forecasts. However, the reconciled forecasts achieve better interpretability than the base forecasts. Akin to the point forecast results, in some birthplaces, there is a very small population with relatively large variability in the number of newborns over the census period. In contrast, in all areas, including those in remote Australia, have population gradually increased with time and slowly varied fertility rates between 1981 and 2011. Consequently, our forecasting method can extract common features of fertility curves from all areas than from all birthplaces, resulting in more accurate interval forecasts.

\section{Conclusion}\label{sec:7}

In this paper, we adapt the hierarchical model and the grouped multivariate functional time series method to forecast Australia's national and sub-national age-specific fertility rates. We highlight the performances of hierarchical arrangements based on the two disaggregation factors, namely the geographical region of residence and birthplace of women, in producing point and interval forecasts of fertility rates. To ensure that independent forecasts at various levels of the hierarchical model consistently sum up to the forecasts at the respective national level, we consider and compare three forecast reconciliation methods. 

Illustrated by the empirical studies on Australian fertility from 1981 to 2011, we found that disaggregating the national fertility rates according to the region of residence of women produces more accurate sub-national forecasts than disaggregation by birthplace. The grouped multivariate functional time series method captured the common features of the groups of fertility curves and produced more accurate forecasts than the conventional univariate functional time series method. Reconciling the independent forecasts by the bottom-top, optimal combination, and trace minimization methods improved point and interval forecast accuracy and interpretability. Among the three considered forecast reconciliation methods, the trace minimization method generally produced the most accurate forecasts for the Australian fertility data.

The superiority of the proposed fertility forecasting methods over the independent functional data method is manifested by multiple sub-national populations with large fertility variability over age and year. For example, age-specific fertility rates of foreign-born women in the United States between 1990--2019 are higher than those of native-born women for all age groups, with significant changes of variations by age for women of all origins \citep{CB}. Because of the better forecasting performance of the multivariate functional time series methods over the independent functional data method, government agencies and statistical bureaus may consider the proposed methods for short-term demographic forecasting. For long-term forecast horizons, assumptions on possible future fertility trends are as important as time series extrapolation methods since the extracted features in the past are not necessarily informative in the longer-term future.

There are at least four ways in which the present paper can be further extended. In the current study, fertility data are grouped according to women's regions of residence or birthplace. The observed fertility curves can still be heterogeneous within some regions or birthplaces. A classification algorithm of \cite{TSY22} may be embedded into statistical models to improve forecast accuracy. In addition, the grouped multivariate functional time series method is not capable of directly capturing the cohort effects of age-specific fertility, that is, the effect of year-of-birth on fertility rates. It is possible to add a cohort factor to the proposed method to capture fertility patterns specific to particular cohorts of women. In analogy to the cohort factor considered by \cite{RH06}, a component depending on $t-x$ may be included after the linear combination of principal components to capture any left out cohort information. Third, the summing matrix is deterministic, a possible research direction is to extend it to time-varying summing matrix, see the Appendix. Finally, forecast combination is known to improve forecast accuracy. In future research, it is possible to combine forecasts from different reconciliation methods to achieve improved accuracy.

\newpage
\begin{appendices} 
\section*{Appendix: \ MinT reconciliation with time-varying summing matrices} \label{appendix}
\setcounter{equation}{0}
\renewcommand{\theequation}{A.\arabic{equation}}

We demonstrate that the summing matrix $\bm{S}_t$ can be used in the reconciliation. We consider a functional time series $\{\bm{f}_t: t \in \mathbb{Z}\}$, where $\bm{f}_t$ are square-integrable functions in the Hilbert space. Denote the eigenfunctions of the long-run covariance function $\bm{C}$ of time series of functions $\left\{\bm{f}_1, \ldots, \bm{f}_t\right\}$ as $\phi_j$ with $j = 1, 2, \ldots$, and denote the corresponding eigenvalues as $\lambda_j$. 
\begin{definition} \label{def_1}
The first $K$ eigenvalues are nonzero and satisfy that $\lambda_1 \geq \lambda_2 \geq \ldots \geq \lambda_K \geq \lambda_{K+1}$. The eigenfunctions related to these largest eigenvalues, i.e., $\bm{\Phi} = \left\{\phi_1, \ldots, \phi_K \right\}$ are the dynamic functional principal components (FPCs) of time series $\{\bm{f}_t; t = 1, 2, \ldots\}$.
\end{definition}
For functional time series satisfying Definition~\ref{def_1}, we need the following assumptions to complete the proof:
\begin{assumption} \label{asp_2}
The empirical functional principal components $\widehat{\bm{\Phi}} = \{\widehat{\phi}_1, \ldots \widehat{\phi}_K\}$ estimated via an eigendecomposition of the sample long-run covariance function span a space that is consistent with the space spanned by $\bm{\Phi} = \left\{\phi_1, \ldots, \phi_K \right\}$.
\end{assumption}

\begin{remark}
Definition~\ref{def_1} states the existence of the dynamic FPCs of a time series $\{\bm{f}_t: t \in \mathbb{Z}\}$ in the Hilbert space. It is common to assume a set of FPCs independent of time for stationary functional time series, e.g., see Definition 2 of \cite{HKH15}. For nonstationary time series, the FPCs are not consistently estimated via the decomposition of an empirical long-run covariance function. However, the span of the basis functions is still consistent \citep{Liebl2013}, and hence is reasonable to adopt Assumption~\ref{asp_2} in the forecast of linear combinations of the FPCs. Our approach is similar to that of \cite{SY19}, \cite{SH17}, \cite{LB09} and others, regarding the use of an FPC analysis for nonstationary data.
\end{remark}

\begin{assumption} \label{asp_3}
The time series of elements in summing the matrices $\{\bm{S}_t: t \in \mathbb{Z}\}$ are stationary with $\mathbb{E}\left[\bm{S}_{t+h}|\I_t\right] = \mathbb{E}\left[\bm{S}_t\right] = \widetilde{S}$. In addition, $\bm{S}_t$ closely follows its mean value in the considered period, resulting in $\bm{S}_{t+h} \approx \mathbb{E}\left[\bm{S}_{t+h}|\I_t\right]$ for a small $h \in \mathbb{Z}^{+}$.
\end{assumption}

\begin{remark}
Let $p_t = \E_{X\ast W,t}/\E_{Y\ast W',t} $ denote a nonzero ratio in the summing matrix $\bm{S_t}$, where $\E_{X\ast W,t}$ and $\E_{X\ast W',t}$ are the ERP exposures in the same year (e.g., $\E_{X\ast W,t} = \E_{\text{A1} \ast j, t}$ representing the number of women in Sydney born in the $j$\textsuperscript{th} birthplace in year $t$, whereas $\E_{X\ast W',t} = \E_{\text{R1} \ast j, t}$ representing the number of women in Region 1 born in the $j$\textsuperscript{th} birthplace in year $t$). Assumption~\ref{asp_3} specifies that $\{p_t: t \in \mathbb{Z} \}$ is a stationary time series. Further, all large and stochastic $p_t$'s in the summing matrices relate to capital cities in Australia or major COB birthplaces, with the remaining stochastic exposure ratios close to $0$ in values. It is seen that women populations in NSW, Victoria, Queensland, Western Australia, and South Australia gradually increased between 1981--2011 at similar speeds, displaying clear linear trends of ERP over the decades \citep{ABSpopulation_chart}. As a result, the large nonzero and stochastic ratios in the summing matrices closely track their time series mean values. Hence, for simplicity we assume that $\mathbb{E}\left[\bm{S}_t\right] = \widetilde{S}$, and $\bm{S}_{t+h} \approx \mathbb{E}\left[\bm{S}_{t+h}|\I_t\right]$ for a small $h \in \mathbb{Z}^{+}$. In practice, constant summing matrices $\bm{S}$ are often considered for reconciliation of subnational time series in Australia, \citep[see, e.g.,][]{WAH19, SY19}, which can be viewed as a special case of our $\bm{S}_t$ settings.
\end{remark}

\begin{assumption} \label{asp_4}
The base level point forecasts $\widehat{\bm{B}}_{t+h|t} = \mathbb{E}\left[\bm{B}_{t+h}|\I_t, \widehat{\bm{\Phi}} \right] \approx \mathbb{E}\left[\bm{B}_{t+h}|\I_t \right]$.
\end{assumption}

\begin{remark}
Assumption~\ref{asp_2} indicate that empirical FPCs $\widehat{\bm{\Phi}}$ estimated from a sample of functions $\{\bm{f}_1, \ldots, \bm{f}_t\}$ are close approximations to the population FPCs $\bm{\Phi}$. Based on the empirical FPCs $\widehat{\bm{\Phi}}$, the point forecasts at the most disaggregated levels obtained by a univariate time series forecasting method of \cite{HS09} can be considered as close approximations to the conditional mean of $\bm{B}_{t+h}$ given the information up to time $t$.
\end{remark}

\begin{assumption} \label{asp_5}
Given information up to time $t$, $\mathbb{E}\left[\bm{S}_{t+h} \bm{B}_{t+h}|\I_t\right] \approx \mathbb{E}\left[\bm{S}_{t+h} |\I_t\right] \mathbb{E}\left[\bm{B}_{t+h}|\I_t\right]$.
\end{assumption}

\begin{remark}  
Assumption~\ref{asp_5} is a relatively strong assumption. However, it may be justified in our application from the following perspective. The conditional variance between $\bm{S}_{t+h}$ and $\bm{B}_{t+h}$ can be expressed as $\text{Cov}\left[\bm{S}_{t+h},\bm{B}_{t+h} |\I_t\right] = \mathbb{E}\left\{\left(\bm{S}_{t+h} - \mathbb{E}\left[\bm{S}_{t+h} |\I_t\right]\right)\left(\bm{B}_{t+h} - \mathbb{E}\left[\bm{B}_{t+h}|\I_t\right]\right) |\I_t \right\}$. According to Assumption~\ref{asp_3}, we then have $\text{Cov}\left[\bm{S}_{t+h},\bm{B}_{t+h} |\I_t\right] \approx 0$, leading to $\mathbb{E}\left[\bm{S}_{t+h} \bm{B}_{t+h}|\I_t\right] \approx \mathbb{E}\left[\bm{S}_{t+h} |\I_t\right] \mathbb{E}\left[\bm{B}_{t+h}|\I_t\right]$.
\end{remark}

For any $t = 1,2,\ldots$, let the summing matrix $\bm{S}_t$ be a matrix of $t$-dependent random variables. Let $\bm{B}_t$ be a vector of observations at the most disaggregated level. Let $\bm{f}_t$ denote a vector of all series in the hierarchy. We leave out subscripts denoting country of birth (COB) or geographical locations for the moment. Following reconciliation equations specified in Section~\ref{sec:4}, we have
\begin{equation*}
\bm{f}_t = \bm{S}_t \bm{B}_t.
\end{equation*}

Let $\widehat{\bm{f}}_{t+h|t}$ denote a vector of $h$-step-ahead base forecasts obtained on training data up to time~$t$. Denote the information set up to time $t$ containing all historical observations including fertility rates and exposures as $\I_t = \{\bm{f}_1, \ldots, \bm{f}_t; \bm{E}_1, \ldots \bm{E}_t\}$, where $\bm{E}_t$ denotes the ERP exposures across all levels at time $t$.  By definition, the $h$-step-ahead forecast error is given by $\widehat{\bm{e}}_{t+h|t} = \bm{f}_{t+h} - \widehat{\bm{f}}_{t+h|t}$. Conditioning on observations up to time $t$, the $h$-step-ahead forecasts are unbiased, i.e., $\mathbb{E}\left[\widehat{\bm{e}}_{t+h|t}|\I_t\right] = \mathbb{E}\left[\bm{f}_{t+h}|\I_t\right] - \mathbb{E}\left[\widehat{\bm{f}}_{t+h|t}| \I_t\right] = 0$. Taking conditional expectation of the product of forecasts at the base level and the summing matrix yields
\begin{equation*}
\mathbb{E}\left[\widehat{\bm{f}}_{t+h|t}|\I_t\right]  = \mathbb{E}\left[\widehat{\bm{S}}_{t+h|t} \widehat{\bm{B}}_{t+h|t} | \I_t\right].
\end{equation*}

For any particular $t$ and $h$, we aim at finding a projection matrix $\bm{P}_{t+h}$ that is non-stochastic and satisfies $\bm{P}_{t+h} \widehat{\bm{S}}_{t+h|t} = \bm{I}$. With an appropriate projection matrix $\bm{P}_{t+h}$, the reconciled forecasts across all levels in the hierarchy can be defined as
\begin{equation*}
\widetilde{\bm{f}}_{t+h|t} = \widehat{\bm{S}}_{t+h|t} \bm{P}_{t+h} \widehat{\bm{f}}_{t+h|t}.
\end{equation*}
The reconciled forecasts will be unbiased if 
\begin{equation*}
\mathbb{E}\left[\widetilde{\bm{f}}_{t+h|t}|\I_t\right] = \mathbb{E}\left[\widehat{\bm{S}}_{t+h|t}\bm{P}_{t+h}\widehat{\bm{S}}_{t+h|t}\widehat{\bm{B}}_{t+h|t}|\I_t\right] = \mathbb{E}\left[\bm{f}_{t+h}|\I_t\right].
\end{equation*}

On page 15 of Section~\ref{sec:3}, we introduced the unbiased base level point forecasts $\widehat{\bm{B}}_{t+h|t} = \mathbb{E}\left[\bm{B}_{t+h|t}| \I_t, \widehat{\bm{\Phi}}\right]$, where $\widehat{\bm{\Phi}}$ is the empirical functional principal components estimated from the sample $\{\bm{f}_1, \ldots, \bm{f}_t\}$. We use an ARIMA model paired with the mean squared error loss function to obtain the $h$-step-ahead forecast of the summing matrix. According to Assumption~\ref{asp_3}, $\widehat{\bm{S}}_{t+h|t} = \mathbb{E}\left[\bm{S}_{n+h} | \I_n \right] = \widetilde{\bm{S}} $. We can then show that
\begin{align}
  \mathbb{E}\left[\widehat{\bm{S}}_{t+h|t}\bm{P}_{t+h}\widehat{\bm{S}}_{t+h|t}\widehat{\bm{B}}_{t+h|t}|\I_t\right] & = \mathbb{E}\left[\widehat{\bm{S}}_{t+h|t}\bm{P}_{t+h}\widehat{\bm{S}}_{t+h|t}\widehat{\bm{B}}_{t+h|t}|\I_t\right] \nonumber \\
  & = \mathbb{E}\left[\widehat{\bm{S}}_{t+h|t}\widehat{\bm{B}}_{t+h|t}|\I_t\right] \nonumber \\
  & = \mathbb{E}\left\{\mathbb{E}\left[\widehat{\bm{S}}_{t+h|t}|\I_t\right] \mathbb{E}\left[\bm{B}_{t+h}|\I_t, \widehat{\bm{\Phi}} \right] | \I_t\right\} \nonumber \\
  & = \mathbb{E}\left[\bm{S}_{t+h}|\I_t\right] \mathbb{E}\left[\bm{B}_{t+h}|\I_t, \widehat{\bm{\Phi}} \right] \nonumber \\
  & = \widetilde{\bm{S}}  \mathbb{E}\left[\bm{B}_{t+h}|\I_t, \widehat{\bm{\Phi}} \right] . \label{eq_a1}
\end{align}

We can also show that
\begin{align}
  \mathbb{E}\left[\widehat{\bm{S}}_{t+h|t}|\I_t\right] \mathbb{E}\left[ \widehat{\bm{B}}_{t+h|t}|\I_t\right] & = \mathbb{E}\left\{\mathbb{E}\left[\bm{S}_{t+h}|\I_t\right]| \I_t\right\} \mathbb{E} \left\{ \mathbb{E}\left[\bm{B}_{t+h}|\I_t, \widehat{\bm{\Phi}} \right] | \I_t \right\} \nonumber \\
  & = \mathbb{E}\left[\bm{S}_{t+h}|\I_t\right] \mathbb{E}\left[\bm{B}_{t+h}|\I_t \right] \nonumber \\
  & = \widetilde{\bm{S}} \mathbb{E}\left[\bm{B}_{t+h}|\I_t \right], \label{eq_a2}
\end{align}
where the last equality has been derived using the Tower Property of conditional expectation \citep[see, e.g., page 88 in Chapter 9 of][]{Williams1991}. Under Assumption~\ref{asp_4}, results of \eqref{eq_a1} and~\eqref{eq_a2} indicate that
\begin{equation}
\mathbb{E}\left[\widehat{\bm{S}}_{t+h|t}\widehat{\bm{B}}_{t+h|t}|\I_t\right] \approx \mathbb{E}\left[\widehat{\bm{S}}_{t+h|t}|\I_t\right] \mathbb{E}\left[ \widehat{\bm{B}}_{t+h|t}|\I_t\right]. \label{eq_a3}
\end{equation}

Next, we consider forecast errors after reconciliation, namely $\widetilde{\bm{e}}_{t+h} = \bm{f}_{t+h} - \widetilde{\bm{f}}_{t+h|t}$. It is easy to show that
\begin{align}
\widetilde{\bm{e}}_{t+h} & = \bm{f}_{t+h} - \widetilde{\bm{f}}_{t+h|t} \nonumber \\
& = \bm{f}_{t+h} - \widehat{\bm{S}}_{t+h|t} \bm{P}_{t+h} \widehat{\bm{f}}_{t+h|t} \nonumber \\
& = \bm{f}_{t+h} - \widehat{\bm{S}}_{t+h|t} \bm{P}_{t+h} \left(\bm{f}_{t+h} - \widehat{\bm{e}}_{t+h|t}\right) \nonumber \\
& = \left(\bm{I} - \widehat{\bm{S}}_{t+h|t} \bm{P}_{t+h}  \right)\bm{f}_{t+h} + \widehat{\bm{S}}_{t+h|t} \bm{P}_{t+h}\widehat{\bm{e}}_{t+h|t} \nonumber \\
& = \left(\bm{I} - \widehat{\bm{S}}_{t+h|t} \bm{P}_{t+h}  \right) \left(\bm{f}_{t+h} - \widehat{\bm{S}}_{t+h|h}\bm{B}_{t+h} + \widehat{\bm{S}}_{t+h|h}\bm{B}_{t+h} \right) \nonumber \\
& \quad + \widehat{\bm{S}}_{t+h|t} \bm{P}_{t+h}\widehat{\bm{e}}_{t+h|t} \nonumber \\
& = \widehat{\bm{S}}_{t+h|t}\left(\bm{I} - \bm{P}_{t+h}\widehat{\bm{S}}_{t+h|t} \right) \bm{B}_{t+h} + \left(\bm{I} - \widehat{S}_{t+h|t}\bm{P}_{t+h}\right)\left(\bm{S}_{t+h} - \widehat{S}_{t+h|h}\right)\bm{B}_{t+h} \nonumber  \\
& \quad + \widehat{\bm{S}}_{t+h|t} \bm{P}_{t+h}\widehat{\bm{e}}_{t+h|t} \nonumber  \\
& = \left(\bm{I} - \widehat{S}_{t+h|t}\bm{P}_{t+h}\right)\left(\bm{S}_{t+h} - \widehat{S}_{t+h|h}\right)\bm{B}_{t+h} + \widehat{\bm{S}}_{t+h|t} \bm{P}_{t+h}\widehat{\bm{e}}_{t+h|t}, \label{eq_a4}
\end{align}
where we have used the fact that $\bm{P}_{t+h}\widehat{\bm{S}}_{t+h|h} = \bm{I}$. By Assumptions~\ref{asp_3} and~\ref{asp_5}, taking conditional expectation of \eqref{eq_a4} gives
\begin{align*}
\mathbb{E}\left[\widetilde{\bm{e}}_{t+h}| \I_t\right] \approx \mathbb{E}\left[\widehat{\bm{S}}_{t+h} \bm{P}\widehat{\bm{e}}_{t+h}|\I_t\right].
\end{align*}
We then have the conditional variance of forecast errors after reconciliation given by
\begin{equation*}
\text{var}\left[\widetilde{\bm{e}}_{t+h}|\I_t\right] \approx \widetilde{\bm{S}}\bm{P}\bm{W_h}\bm{P}^{\top}\widetilde{\bm{S}}^{\top},
\end{equation*}
where $\bm{W}_h = \mathbb{E}\left[\widehat{\bm{e}}_{t+h}\widehat{\bm{e}}_{t+h}^{\top}|\I_t\right]$ and $\widetilde{\bm{S}} = \mathbb{E}[\bm{S}_t|\I_t]  = \mathbb{E}\left[\bm{S}_t\right]$. Hence, the optimal projection matrix should minimize $\text{tr}\left[\widetilde{\bm{S}}\bm{P}\bm{W_h}\bm{P}^{\top}\widetilde{\bm{S}}^{\top}\right]$ subject to $\bm{P}_{t+h}\widetilde{\bm{S}} = \bm{I}$. The optimal projection matrix is given by
\begin{align*}
\bm{P}_{t+h} = \left(\widetilde{\bm{S}}^{\top}{\bm{W}}_h^{-1}\widetilde{\bm{S}}\right)^{-1}\widetilde{\bm{S}}^{\top}{\bm{W}}_h^{-1}.
\end{align*}

\end{appendices}

\newpage
\bibliographystyle{agsm}
\bibliography{Australian_fertility.bib}

\end{document}